\documentclass[twocolumn]{aastex62}
\usepackage{amsmath}
\usepackage{ amssymb }
\usepackage{float}

\newcommand{\mgfe}[0]{[{\rm Mg/Fe}]} 
\newcommand{\Acc}{A_{\rm cc}}
\newcommand{\AIa}{A_{\rm Ia}} 
\newcommand{\RIa}{R_{\rm Ia}^X}
\newcommand{\aIa}{\alpha_{\rm Ia}}
\newcommand{\acc}{\alpha_{\rm cc}}

\newcommand{\femg}{[{\rm Fe}/{\rm Mg}]} 
\newcommand{\xmg}{[{\rm X}/{\rm Mg}]} 
\newcommand{\xfe}{[{\rm X}/{\rm Fe}]} 
\newcommand{\mgh}{[{\rm Mg}/{\rm H}]}
\newcommand{\feh}[0]{[{\rm Fe/H}]} 
\newcommand{\pIa}{p_{\rm Ia}^{\rm X}}
\newcommand{\pcc}{p_{\rm cc}^{\rm X}}
\newcommand{\pIasun}{p_{\rm Ia, \odot}^{\rm X}}
\newcommand{\pccsun}{p_{\rm cc, \odot}^{\rm X}}
\newcommand{\fcc}{f_{\rm cc}}
\newcommand{\logg}{\log(g)}
\newcommand{\teff}{T_{\rm eff}}
\newcommand{\kpc}{\rm \; kpc}
\newcommand{\kel}{\rm \; K}
\newcommand{\msun}{\rm M_{\odot}}
\newcommand{\rgc}{R_{\rm GC}}
\newcommand{\msunvice}{\rm M_{\odot}/ \rm M_{\odot formed}}

\defcitealias{weinberg}{W19}
\defcitealias{jonnson20}{J20}
\defcitealias{andrews17}{AWSJ17}
\defcitealias{CL13}{CL13}
\defcitealias{LC18}{LC18}
\defcitealias{griffith19}{GJW}

\newcommand{\add}[1]{\textcolor{black}{#1}}

\usepackage[section]{placeins}

\received{2020 September 10}
\revised{2020 December 21}
\accepted{2020 December 22}

\shorttitle{Abundance Ratio Trends of the Milky Way Bulge}
\shortauthors{Griffith et al.}

\begin{document}

\title{The Similarity of Abundance Ratio Trends and Nucleosynthetic Patterns in the Milky Way Disk and Bulge}

\correspondingauthor{Emily Griffith}
\email{griffith.802@osu.edu}

\author[0000-0001-9345-9977]{Emily Griffith}
\affiliation{The Department of Astronomy and Center of Cosmology and AstroParticle Physics, The Ohio State University, Columbus, OH 43210, USA}

\author{David H. Weinberg}
\affiliation{The Department of Astronomy and Center of Cosmology and AstroParticle Physics, The Ohio State University, Columbus, OH 43210, USA}

\author{Jennifer A. Johnson}
\affiliation{The Department of Astronomy and Center of Cosmology and AstroParticle Physics, The Ohio State University, Columbus, OH 43210, USA}


\author{Rachael Beaton}
\affiliation{The Carnegie Observatories, 813 Santa Barbara Street, Pasadena, CA 91101, USA}

\author{D. A. Garc\'{i}a-Hern\'{a}ndez}
\affiliation{Instituto de Astrof\'{i}sica de Canarias (IAC), E-38205 La Laguna, Tenerife, Spain}
\affiliation{Universidad de La Laguna (ULL), Departamento de Astrof\'{i}sica, E-38206 La Laguna, Tenerife, Spain}

\author{Sten Hasselquist}
\altaffiliation{NSF Astronomy and Astrophysics Postdoctoral Fellow}
\affiliation{Department of Physics \& Astronomy, University of Utah, Salt Lake City, UT, 84112, USA}

\author{Jon Holtzman}
\affiliation{Department of Astronomy, New Mexico State University, Las Cruces, NM 88003, USA}

\author{James W. Johnson}
\affiliation{The Department of Astronomy and Center of Cosmology and AstroParticle Physics, The Ohio State University, Columbus, OH 43210, USA}

\author[0000-0002-4912-8609]{Henrik J\"onsson}
\affiliation{Materials Science and Applied Mathematics, Malm\"o University, SE-205 06 Malm\"o, Sweden}
\affiliation{Lund Observatory, Department of Astronomy and Theoretical Physics, Lund University, Box 43, SE-22100 Lund, Sweden}

\author{Richard R. Lane}
\affiliation{Instituto de Astronom\'{i}a y Ciencias Planetarias de Atacama, Universidad de Atacama, Copayapu 485, Copiap\'{o}, Chile}

\author{David M. Nataf}
\affiliation{Center for Astrophysical Sciences and Department of Physics and Astronomy,
The Johns Hopkins University,
Baltimore, MD 21218}

\author[0000-0002-1379-4204]{Alexandre Roman-Lopes}
\affiliation{Departamento de Astronom\'ia, Universidad de La Serena - Av. Juan Cisternas, 1200 North, La Serena, Chile}


\begin{abstract}
We compare abundance ratio trends in a sample of $\sim 11,000$ Milky Way bulge stars ($\rgc < 3\kpc$) from the Apache Point Observatory Galactic Evolution Experiment (APOGEE) to those of APOGEE stars in the Galactic disk ($5 \kpc < \rgc < 11 \kpc$). We divide each sample into low-Ia (high-$\mgfe$) and high-Ia (low-$\mgfe$) populations, and in each population we examine the median trends of $\xmg$ vs. $\mgh$ for elements X = Fe, O, Na, Al, Si, P, S, K, Ca, V, Cr, Mn, Co, Ni, Cu, and Ce. To remove small systematic trends of APOGEE abundances with stellar $\logg$, we resample the disk stars to match the $\logg$ distributions of the bulge data. After doing so, we find nearly identical median trends for low-Ia disk and bulge stars for all elements. High-Ia trends are similar for most elements, with noticeable (0.05-0.1 dex) differences for Mn, Na, and Co. The close agreement of abundance trends (with typical differences $\lesssim 0.03$ dex) implies that similar nucleosynthetic processes enriched bulge and disk stars despite the different star formation histories and physical conditions of these regions. For example, we infer that differences in the high mass slope of the stellar initial mass function (IMF) between disk and bulge must have been $\lesssim 0.30$. This agreement, and the generally small scatter about the median sequences, means that one can predict all of a bulge star's APOGEE abundances with good accuracy knowing only its measured $\mgfe$ and $\mgh$ and the observed trends of {\it disk} stars. 
\end{abstract}


\section{Introduction} \label{sec:intro}

To paint a complete picture of our Galactic enrichment history, we need to study the chemical fingerprints of stars from the outer edges to the inner depths of the Milky Way.
Large scale Galactic surveys like GALAH\footnote{GALactic Archaeology with HERMES} \citep{desilva2015, martell2017}, Gaia-ESO \citep{gilmore2012}, and APOGEE\footnote{APOGEE = Apache Point Observatory Galactic Evolution Experiment, currently a part of the Sloan Digital Sky Survey IV (SDSS-IV)} \citep{blanton2017, majewski17} give us the power to study stellar populations throughout vast regions of the Milky Way. Each provides a glimpse of the past and unveils the chemical makeup of the interstellar medium (ISM) in which the stars were born. 
Using APOGEE data, \citet{nidever15} and \citet{hayden2015} showed that stellar populations follow the same trends in [$\alpha$/Fe] vs. $
\feh$ throughout the Milky Way disk, though the distribution of stars along these tracks depends strongly on galactocentric radius $\rgc$ and midplane distance $|Z|$. Extending these results, \citet[][hereafter W19]{weinberg} found that the median trends of $\xmg$ vs. $\mgh$ for all APOGEE elements are independent of Galactic location, suggesting universality in the nucleosynthetic processes that determine these abundance ratios. In this work, we aim to extend their conclusions to the Galactic bulge by leveraging the extensive bulge coverage of APOGEE DR16 to study its stellar chemical abundances. 

In addition to these large scale programs, many groups have observed smaller ($\lesssim 100$ stars) samples of the inner Galaxy and debated their chemical similarity to the disk. Works such as \citet{mcwilliam1994} (12 giants), \citet{cunha2006} (7 giants), \citet{fulbright2007} (27 giants), and \citet{lecureur2007} (53 giants) found enhancements in the bulge $\alpha$-element abundances relative to disk samples\footnote{$\alpha$ elements such as O, Mg, and Si are produced mainly by core collapse supernovae while iron peak elements in stars with solar [$\alpha$/Fe]=0 have roughly equal contributions from core collapse and Type Ia supernovae.}. However, subsequent work by \citet{melendez2008} (20 giants) found bulge [O/Fe] values in line with their thick-disk sample. More recently the bulge/disk chemical similarities/differences have been examined by studies of both giant \citep[e.g.][]{johnson2014, jonsson2017, lomaeva2019, forsberg2019, duong2019A, duong2019B} and dwarf \citep[e.g.][]{bensby2013, bensby2017} stars. These works span $\alpha$-elements to neutron capture elements and include all those observed by APOGEE. Within these works, small differences between the bulge and thick disk [X/Fe] vs. [Fe/H] median abundance trends or abundance distributions are found (see above citations). We note that the collective studies do not come to a consensus on bulge/thick disk abundance differences.

Large scale surveys such as APOGEE can provide more substantial bulge coverage than these smaller studies. While the number of APOGEE bulge observations drastically increased with the inclusion of DR16 data \citep{sdss16}, the bulge has been studied in prior data releases as well. \citet{zasowski2019} present [X/Fe] vs. [Fe/H] abundance distributions and median trends of 4000 bulge stars ($\rgc < 4 \kpc$) using DR14. They compare bulge median trends with a solar radius population, finding good agreement with their high-$\alpha$ population but not with their low-$\alpha$ stars, especially at low $\feh$. 
APOGEE observations of 424 bulge stars through Baade's window agree with prior bulge works, suggesting that APOGEE DR13 data do not suffer systematic biases in the bulge \citep{schultheis2017}. 

All of the prior studies cited here present bulge abundances in [X/Fe] vs. [Fe/H] space and generally treat the bulge as a single population, comparable to the thick disk. It is well known that stellar populations at the solar annulus show two distinct trends in [$\alpha$/Fe] vs. $\feh$ space \citep{fuhrmann1998, prochaska2000, bensby2003}. These are commonly referred to as the high-$\alpha$ and low-$\alpha$ populations, though in this paper we will refer to them as low-Ia and high-Ia, respectively, since the pre-physical differences between them arise from the Type-Ia supernova contribution to their iron abundances. \add{While some studies \cite[e.g.][]{jonsson2017, zasowski2019} do identify high and low-Ia stars,} the bimodality of [$\alpha$/Fe] is less evident in the bulge, so most studies have treated the bulge as a single evolutionary sequence. 
\add{When the number of stars is small, the two populations blur together.}
However a recent analysis of APOGEE DR14 bulge ($R<3.5 \kpc$) stars by \citet{rojas} found the [Mg/Fe] distribution to be bimodal for stars near solar $\feh$, with a distinct low-[Mg/Fe] (high-Ia) sequence that had been previously categorized as a continuation of the low-Ia population. Other recent works with APOGEE DR16 data confirm that the bulge exhibits a disk-like double sequence \citep[e.g.][]{bovy2019, queiroz20, lian2020}.

Stellar populations that span high to low-$\mgfe$ can be decomposed into high-Ia and low-Ia components, as done in \citetalias{weinberg}. Unlike prior bulge studies, we can now leverage the large bulge population observed by APOGEE DR16 to separately study the high-Ia and low-Ia populations of the inner Galaxy. When viewed in [X/Mg] vs. [Mg/H] space, these two populations inform us on the relative contribution of core collapse supernovae (CCSN) and Type-Ia supernovae (SNIa). Elements dominantly produced by prompt CCSN enrichment (e.g. O, Mg, Si, Ca) will display high-Ia and low-Ia sequences with little separation, while elements with an increased SNIa contribution (e.g. Fe, Ni, Mn) will show a larger sequence separation. The two-process decomposition model developed by \citetalias{weinberg} allows us to quantify the relative prompt and delayed contribution to each element. By studying the abundances in Mg space rather than Fe space, we can conduct a clearer comparison of the nucleosynthetic contributions to the bulge and disk than has been done before, since Mg comes almost entirely from CCSN. .

Using APOGEE DR14 disk stars, \citetalias{weinberg} showed that the median trends of [X/Mg] vs. [Mg/H] for the low-Ia and high-Ia populations remained constant throughout the disk for O, Na, Mg, Al, Si, P, S, K, Ca, V, Cr, Mn, Fe, Co, and Ni. \citet[][hereafter GJW]{griffith19} expanded this work to many new elements in GALAH and drew similar conclusions. 
In this paper we use APOGEE DR16 data to carry out a similar analysis of the bulge. Previous bulge studies have focused mostly on the distribution of stars in [Fe/H], [$\alpha$/Fe], or other element ratios. These distributions are sensitive to many aspects of chemical evolution such as star formation history, star formation efficiency, and outflows. Here were focus on the median trends of $\xmg$ vs. $\mgh$, which are sensitive mainly to the relative nucleosynthetic yields of these elements (see \citetalias{weinberg}). Comparing disk and bulge sequences therefore allows us to ask whether the supernovae that enriched the bulge are similar to the supernovae that enriched the disk, even if the relative contributions of CCSN and SNIa are different.

After discussing our data selection in Section~\ref{sec:data}, we present the APOGEE DR16 bulge abundances in Mg and Fe space in Section~\ref{sec:abundnaces}. Here we also construct a comparison disk sample and analyze the similarities and differences between the bulge and disk median high-Ia and low-Ia trends. In Section~\ref{sec:two-proc} we review the two-process model of \citetalias{weinberg}, apply it to the disk trends, and predict the bulge abundances based on the best fit parameters. Section~\ref{sec:IMFs} contains a discussion of potential constrains on the IMF difference between the bulge and the disk. We summarize our work in Section~\ref{sec:summary}.

\section{Data} \label{sec:data}

In this paper, we use data from the sixteenth data release (DR16) of the SDSS/APOGEE survey \citep[][hereafter J20]{sdss16, jonnson20}. DR16 extends APOGEE's Galactic view to the southern hemisphere, as we observe with two nearly identical, 300 fiber APOGEE spectographs \citep{wilson19} on the 2.5 m Sloan Foundation telescope \citep{gunn06} at Apache Point Observatory (APO) in New Mexico and the 2.5 m du Pont telescope \citep{bowen73} at Las Campanas Observatory (LCO) in Chile. The addition of LCO allows APOGEE to observe a greater number of stars in the inner Galaxy, crucial for the work done in this paper. Targeting selection for APOGEE-2 is described by \citet{zasowski17} and is updated in Santana et al. (in prep.). Bulge stars are selected as a part of the main APOGEE red giant sample.
The APOGEE data are reduced as described by \citet{nidever15} and fed into the APOGEE Stellar Parameters and Chemical Abundances Pipeline \citep[ASPCAP;][]{holtzman15, holtzman18, garcia16}. ASPCAP returns the best fit effective temperatures, surface gravities, and elemental abundances employed in this paper. See \citetalias{jonnson20} and references therein for a more detailed description of the DR16 data reduction analysis and validation.

We use data from an internal data release including observations through November 2019 rather than the public DR16. These additional observations have been processed with the same DR16 pipeline, so we will call this data set DR16+\footnote{APOGEE allStar file allStar-r13-l33-58814}. The ~4 months of additional observations add many more bulge fields, increasing the number of stars in our bulge population from 6,978 in DR16 to \add{11,229} in DR16+. The additional data and their ASPCAP analysis will be included in APOGEE DR17 (planned for mid-2021). We have repeated our analysis with the smaller DR16 data set and find very similar results. DR16+ provides more robust median trends for difficult to observe elements (e.g. Na, K, Mn, Cu, Ce) and for the less populated, low-metallicity regime. 

We apply quality cuts to the DR16+ APOGEE catalog to extract stars with reliable, calibrated abundance measurements. We cut all stars with flags set for many bad pixels, bright neighbors, high persistence levels, broad lines, or radial velocity warnings (STARFLAGs 0, 3, 9, 16, and 17) as well as those with expected bad or dubious determinations of $\teff$, $\logg$, metallicity, and $\alpha$ element content (ASPCAPFLAGs 0, 3, 16, 17). Duplicate observations are removed. We require there to be no elemental flags for [Mg/Fe] and [Fe/H] in our main sample, and we further discard stars with flagged abundances in their respective elemental analyses. After removing flagged stars we make a signal-to-noise ratio (SNR) cut of SNR$>100$.  

Finally, we remove the cool stars by requiring $T_{\text{eff}} >3500$ K. For some elements cool stars produce systematic structures in [X/Fe] vs. [Fe/H] space, likely due to a failure in the APSCAP fit. Our cut at 3500 K does not remove all observational artifacts, but it significantly cleans up the sample. A larger discussion of these artifacts can be found in Section~\ref{subsec:artifacts} and in \citetalias{jonnson20}. 

\begin{figure}[]
 \includegraphics[width=\columnwidth]{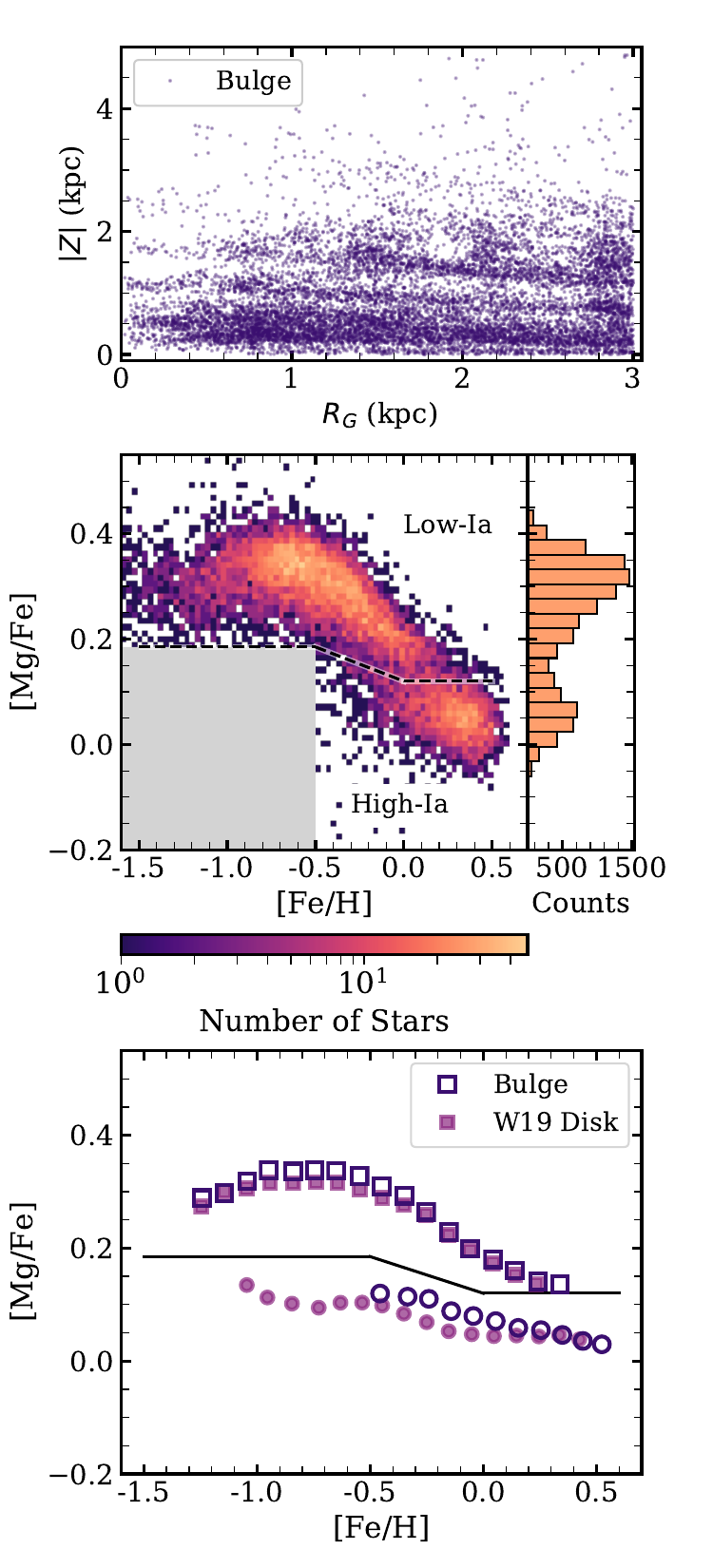}
 \caption{Top: The distribution of our bulge stellar sample in the $|Z|$ vs. $\rgc$ plane. Middle: Density plot of stars in the bulge. The dashed line denotes our division between the high-Ia and low-Ia populations. We exclude high-Ia stars with $\feh < -0.5$ \add{and $\mgfe < -0.185$}, located in the grey region. We include a histogram of our bulge sample's $\mgfe$ distribution in the right hand panel. Bottom: Median trends for the low-Ia (squares) and high-Ia (circles) bulge populations are shown in dark purple. Those for the disk (as defined by \citetalias{weinberg}) are shown in pink. }
 \label{fig:bulge_den}
\end{figure}

Stars that pass these cuts span the Milky Way. While the boundaries between the disk, inner Galaxy, and bulge are debated, we define the Galactic bulge as $\rgc < 3 \kpc$. We apply a height cut of $|Z|<5 \kpc$.
We use spectrophotometric distances from \citet{rojas2020} and take the Galactic center to be at 8.178 kpc \citep{abuter}. To assure that our stars reside in the inner galaxy we remove those with quoted distance errors greater than 10\% of their distance. While we choose to use the \citet{rojas2020} distances, \texttt{astroNN} \citep{bovy2019, leung19} and \texttt{StarHorse} \citep{queiroz20} distances are also publicly available for DR16 data (see the cited papers for more detailed comparison of methods and results). When we repeat our analysis with these two alternative distance sets, we find good agreement between the derived median abundance trends. Our results are therefore independent of the specific distance derivation used. The distribution of our bulge sample in $|Z|$ vs. $\rgc$ is shown in the top panel of Figure~\ref{fig:bulge_den}.

These quality and spatial cuts leave us with a sample of 11,229 stars in the bulge, which have a median SNR of 149. The middle panel of Figure~\ref{fig:bulge_den} shows the density of these stars in $\mgfe$ vs. $\feh$ space and the distribution of their $\mgfe$ values. We define the low-Ia population as stars which satisfy the following set of equations: 
\begin{equation}
    \begin{cases}
    \mgfe \geq 0.185,           & \feh \leq -0.5 \cr
    \mgfe \geq 0.12 - 0.13\feh,    & -0.5< \feh \leq 0 \cr
    \mgfe \geq 0.12,               & \feh > 0. \cr
    \end{cases}
    \label{eq:boundary}
\end{equation}
This dividing line is plotted on Figure~\ref{fig:bulge_den}. We note that our division differs from that of \citetalias{weinberg} at $\feh\leq-0.5$, as we add a plateau to successfully separate the low metallicity stars. We further require high-Ia stars to have $\feh>-0.5$ to avoid contamination with the stars in the low-Ia population that scatter below \add{$\mgfe=-0.185$} at lower metallicities. \add{This exclusion removes 153 stars. Some excluded objects are likely true low-[Fe/H] high-Ia stars, but their low density in abundance space would not meet our criteria for median trend analysis.}

While previous studies of the bulge have seen a single, continuous sequence in the [$\alpha$/Fe] vs. \feh plane \citep[e.g.][]{ness2016}, the histogram in Figure~\ref{fig:bulge_den} strongly suggests the presence of two distinct populations. The distribution of $\mgfe$ has a minimum at $\mgfe \approx 0.13$ with bimodality particularly evident near $\feh \approx 0$, as seen previously by \citet{rojas}. However, compared to the disk (see Figure~\ref{fig:disk_den}), the high-Ia sequence is less evident at $\feh < 0$. The $\mgfe$ distribution is affected by the height distribution of the sample \citep[e.g.][]{bovy2016}. Here, we have not corrected for the latitude sampling bias, so our $\mgfe$ distribution may not be representative of the full bulge population. In their study of the bulge bimodality, \citet{lian2020} do correct this bias. After corrections they find a stronger signal of bimodality than in the parent population. We likely see more low-Ia stars than would be representative of this Galactic region due to the higher latitude observations. For the purposes of this paper, the separation of our sample into two populations is important mainly because it allows us to make accurate comparisons of bulge and disk populations with similar levels of SNIa enrichment. We do not make conclusions about the relative number of stars on the low-Ia or high-Ia sequence, nor the bulge MDF, so we do not discuss the sampling bias further.

Compared to the Galactic disk (\citetalias{weinberg}; Figure 1), the low-Ia sequence of the bulge remains well populated to lower metallicity, though the low-Ia sequences of both the bulge and the disk reach an $\feh$ of $\sim -1.3$. The bottom panel of Figure~\ref{fig:bulge_den} shows the median trends of the high-Ia and low-Ia populations for the bulge and disk. Here, we use the \citetalias{weinberg} disk definition ($3 \kpc < R < 15 \kpc$, $|Z| < 2\kpc$, $1 < \logg < 2$, and $3700 \kel < \teff < 4600 \kel$), but recalculate with DR16+ data and the quality cuts listed above. We will refer to this sample as ``W19 disk'' throughout the paper. The $\mgfe$ vs. $\feh$ median trends for W19 and the bulge agree to within $\sim0.05$ dex at all metallicities. Agreement is further improved by matching the $\logg$ distributions of the disk and bulge samples as discussed below (see Section~\ref{subsec:sample}). We find that the $\mgfe$ downturn, or knee, of both populations occurs at the same $\feh$, in agreement with \citet{zasowski2019}. 

\section{Bulge Stellar Abundances} \label{sec:abundnaces}

\begin{figure*}[!htb]
 \begin{centering}
 \includegraphics[width=\textwidth, angle=0]{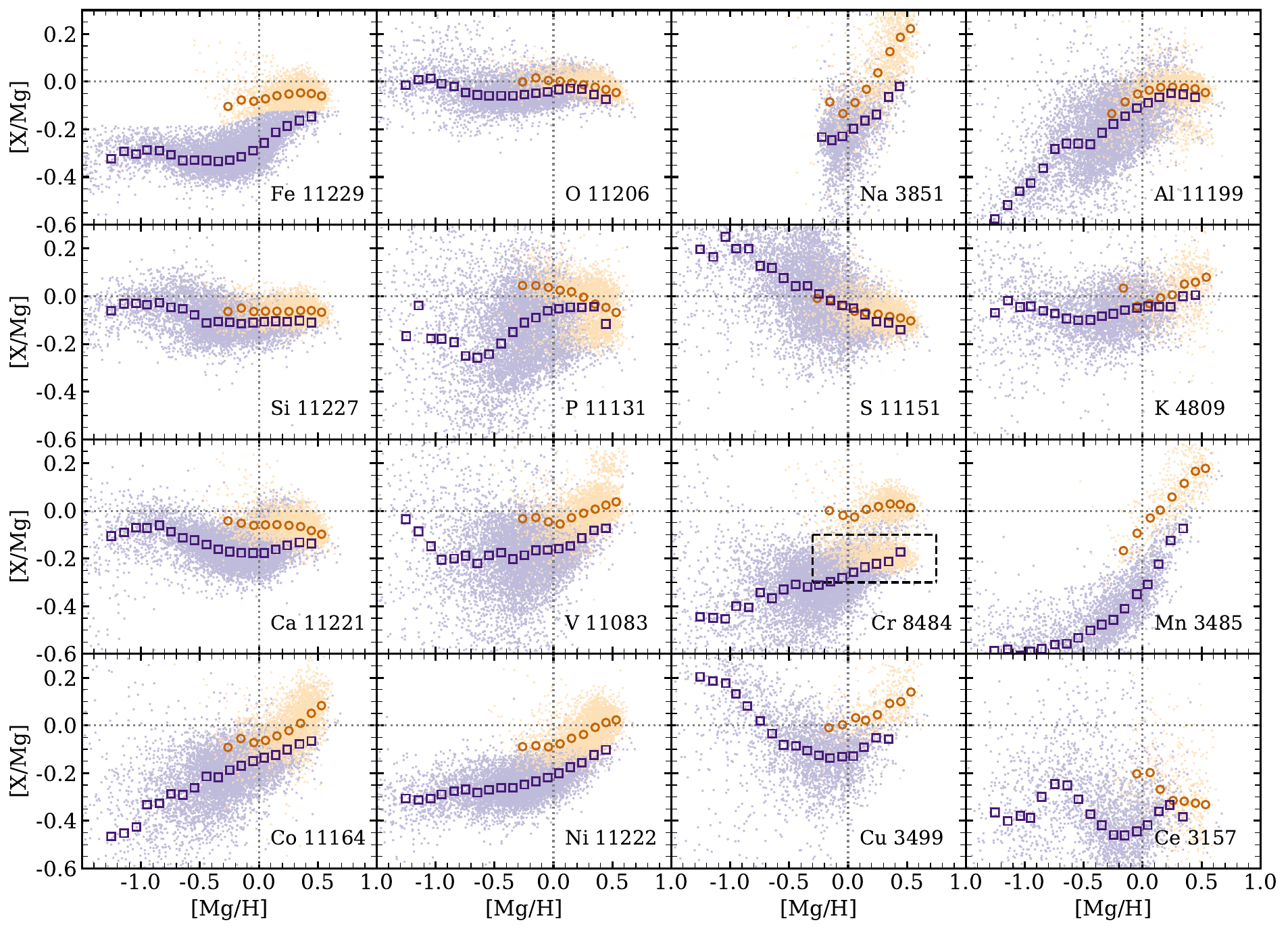}
 \caption{Bulge stellar abundance distributions in [X/Mg] vs. [Mg/H] space for APOGEE elements. High-Ia stars are in orange and low-Ia stars are in purple. The medians of the high/low-Ia populations are over-plotted, where we bin by 0.1 dex and require $> 20$ stars per bin. The dotted lines in each panel denote solar [X/Mg] and [Mg/H]. The number in the bottom right hand corner of each subplot corresponds to the number of stars for each elemental population which pass our cuts. High-Ia stars with Cr abundances within the dashed box are excluded in the calculation of the high-Ia median.}
 \label{fig:XMg}
 \end{centering}
\end{figure*}

In this section, we discuss the median abundance trends of the APOGEE DR16+ bulge stars (see Tables~\ref{tab:medians1} and~\ref{tab:medians2}). Figure~\ref{fig:XMg} plots the APOGEE abundances in $\xmg$ vs. $\mgh$ space for Fe, O, Na, Al, Si, P, S, K, Ca, V, Cr, Mn, Co, Ni, Cu, and Ce along with the median trends for the high-Ia and low-Ia populations. Median values are calculated in bins of width 0.1 dex in $\mgh$ and require $>20$ stars per bin.  \add{APOGEE abundances are calculated with LTE assumptions. See \citet{osorio2020} for a discussion of the potential NLTE effects on Na, Mg, K, and Ca.}

\begin{deluxetable*}{ccccccccc}[ht]
\tablecaption{Median high-Ia (top) and low-Ia (bottom) sequences for APOGEE DR16+ $\alpha$-elements and light-$Z$ elements. We calculate medians in bins with a width of 0.1 dex in [Mg/H], requiring $> 20$ stars per bin. Zero-point shifts discussed in Section~\ref{subsec:fitting_2proc} are included. \label{tab:medians1}}
\tablehead{
\colhead{[Mg/H]} & \colhead{[O/Mg]} & \colhead{[Na/Mg]} & \colhead{[Al/Mg]} & \colhead{[Si/Mg]} & \colhead{[P/Mg]} & \colhead{[S/Mg]} & \colhead{[K/Mg]} & \colhead{[Ca/Mg]} 
}
\startdata
-0.262 & -0.044 & 0.002 & -- & -0.085 & -0.018 & 0.022 & 0.025 & --  \\ 
-0.149 & -0.017 & 0.019 & -0.041 & -0.036 & -0.004 & 0.022 & 0.022 & 0.048  \\ 
-0.043 & -0.022 & 0.008 & -0.090 & -0.003 & -0.018 & 0.014 & -0.006 & -0.028  \\ 
0.056 & -0.012 & 0.005 & -0.044 & 0.013 & -0.017 & 0.002 & -0.029 & -0.017  \\ 
0.153 & 0.001 & -0.003 & 0.012 & 0.025 & -0.017 & -0.004 & -0.034 & 0.007  \\ 
0.255 & 0.008 & -0.010 & 0.081 & 0.026 & -0.018 & -0.027 & -0.041 & 0.020  \\ 
0.354 & 0.013 & -0.020 & 0.170 & 0.023 & -0.014 & -0.056 & -0.051 & 0.065  \\ 
0.445 & 0.010 & -0.030 & 0.230 & 0.019 & -0.014 & -0.070 & -0.057 & 0.073  \\ 
0.532 & -0.000 & -0.043 & 0.266 & 0.003 & -0.021 & -0.091 & -0.069 & 0.094  \\ 
\hline
-1.254 & -0.264 & -0.011 & -- & -0.527 & -0.015 & -0.190 & 0.231 & -0.056  \\ 
-1.145 & -0.232 & 0.011 & -- & -0.468 & 0.015 & -0.062 & 0.199 & -0.005  \\ 
-1.042 & -0.244 & 0.017 & -- & -0.411 & 0.016 & -0.200 & 0.283 & -0.031  \\ 
-0.95 & -0.226 & -0.005 & -- & -0.377 & 0.011 & -0.201 & 0.234 & -0.028  \\ 
-0.844 & -0.229 & -0.017 & -- & -0.314 & 0.019 & -0.216 & 0.233 & -0.047  \\ 
-0.746 & -0.246 & -0.043 & -- & -0.233 & 0.000 & -0.273 & 0.161 & -0.058  \\ 
-0.646 & -0.270 & -0.053 & -- & -0.211 & -0.007 & -0.281 & 0.153 & -0.079  \\ 
-0.547 & -0.269 & -0.056 & -- & -0.211 & -0.032 & -0.266 & 0.110 & -0.086  \\ 
-0.444 & -0.270 & -0.056 & -- & -0.213 & -0.066 & -0.221 & 0.077 & -0.085  \\ 
-0.345 & -0.274 & -0.056 & -- & -0.165 & -0.060 & -0.173 & 0.078 & -0.069  \\ 
-0.248 & -0.269 & -0.051 & -0.188 & -0.128 & -0.063 & -0.134 & 0.044 & -0.060  \\ 
-0.149 & -0.255 & -0.045 & -0.201 & -0.096 & -0.069 & -0.113 & 0.016 & -0.044  \\ 
-0.049 & -0.230 & -0.040 & -0.185 & -0.062 & -0.064 & -0.084 & -0.004 & -0.036  \\ 
0.046 & -0.197 & -0.030 & -0.153 & -0.040 & -0.061 & -0.076 & -0.016 & -0.028  \\ 
0.142 & -0.152 & -0.025 & -0.119 & -0.017 & -0.059 & -0.069 & -0.043 & -0.030  \\ 
0.241 & -0.126 & -0.029 & -0.094 & 0.001 & -0.060 & -0.070 & -0.072 & -0.030  \\ 
0.345 & -0.103 & -0.051 & -0.020 & -0.004 & -0.056 & -0.067 & -0.076 & 0.014  \\ 
0.445 & -0.086 & -0.071 & 0.024 & -0.017 & -0.065 & -0.139 & -0.106 & 0.019  \\ 
\enddata
\end{deluxetable*}

\begin{deluxetable*}{ccccccccc}[ht]
\tablecaption{Same as Table~\ref{tab:medians1}, but for APOGEE DR16+ Fe-peak elements, Cu, and Ce \label{tab:medians2}}
\tablehead{
\colhead{[Mg/H]} & \colhead{[V/Mg]} & \colhead{[Cr/Mg]} & \colhead{[Fe/Mg]} & \colhead{[Mn/Mg]} & \colhead{[Co/Mg]} & \colhead{[Ni/Mg]} & \colhead{[Cu/Mg]} & \colhead{[Ce/Mg]} 
}
\startdata
-0.262 & 0.015 & 0.001 & -- & -- & -0.069 & -0.030 & -- & -- \\ 
-0.149 & 0.005 & 0.006 & 0.040 & -0.200 & -0.032 & -0.026 & -0.045 & -- \\ 
-0.043 & -0.004 & -0.012 & 0.020 & -0.127 & -0.049 & -0.031 & -0.032 & -0.075 \\ 
0.056 & -0.002 & -0.021 & 0.013 & -0.063 & -0.040 & -0.018 & -0.003 & -0.070 \\ 
0.153 & -0.001 & 0.005 & 0.045 & -0.030 & -0.021 & 0.005 & -0.013 & -0.140 \\ 
0.255 & -0.004 & 0.024 & 0.057 & 0.025 & 0.001 & 0.021 & 0.010 & -0.187 \\ 
0.354 & -0.009 & 0.041 & 0.068 & 0.082 & 0.032 & 0.051 & 0.057 & -0.190 \\ 
0.445 & -0.026 & 0.058 & 0.066 & 0.133 & 0.074 & 0.071 & 0.065 & -0.198 \\ 
0.532 & -0.041 & 0.072 & 0.052 & 0.145 & 0.106 & 0.082 & 0.105 & -0.204 \\ 
\hline
-1.254 & -0.049 & -0.001 & -0.406 & -0.620 & -0.443 & -0.247 & 0.169 & -0.237  \\ 
-1.145 & -0.033 & -0.052 & -0.410 & -0.614 & -0.429 & -0.254 & 0.151 & -0.273  \\ 
-1.042 & -0.014 & -0.114 & -0.414 & -0.639 & -0.403 & -0.248 & 0.142 & -0.251  \\ 
-0.95 & -0.015 & -0.171 & -0.361 & -0.622 & -0.309 & -0.230 & 0.097 & -0.259  \\ 
-0.844 & -0.004 & -0.166 & -0.366 & -0.612 & -0.303 & -0.216 & 0.047 & -0.172  \\ 
-0.746 & -0.031 & -0.154 & -0.304 & -0.594 & -0.264 & -0.210 & -0.016 & -0.118  \\ 
-0.646 & -0.056 & -0.187 & -0.328 & -0.590 & -0.268 & -0.223 & -0.069 & -0.123  \\ 
-0.547 & -0.066 & -0.153 & -0.291 & -0.566 & -0.238 & -0.211 & -0.116 & -0.181  \\ 
-0.444 & -0.084 & -0.141 & -0.270 & -0.535 & -0.191 & -0.202 & -0.121 & -0.244  \\ 
-0.345 & -0.104 & -0.168 & -0.280 & -0.510 & -0.194 & -0.203 & -0.140 & -0.291  \\ 
-0.248 & -0.114 & -0.152 & -0.272 & -0.490 & -0.164 & -0.189 & -0.160 & -0.331  \\ 
-0.149 & -0.119 & -0.131 & -0.258 & -0.443 & -0.146 & -0.175 & -0.172 & -0.333  \\ 
-0.049 & -0.120 & -0.130 & -0.242 & -0.383 & -0.126 & -0.160 & -0.166 & -0.316  \\ 
0.046 & -0.120 & -0.125 & -0.219 & -0.342 & -0.112 & -0.141 & -0.164 & -0.290  \\ 
0.142 & -0.105 & -0.113 & -0.197 & -0.256 & -0.101 & -0.116 & -0.126 & -0.233  \\ 
0.241 & -0.088 & -0.080 & -0.183 & -0.157 & -0.078 & -0.097 & -0.087 & -0.206  \\ 
0.345 & -0.075 & -0.048 & -0.174 & -0.107 & -0.055 & -0.065 & -0.092 & -0.255  \\ 
0.445 & -0.080 & -0.039 & -0.134 & -- & -0.042 & -0.043 & -- & --  \\ 
\enddata
\end{deluxetable*}

While APOGEE reports C, N, and Ti for some bulge stars, we do not include them in this paper. We remove the C and N abundances as mixing of processed material in giant stars causes the observed atmospheric abundances to differ from their birth values. We exclude Ti because the Ti I abundances show inaccurate [Ti/Fe] vs. [Fe/H] trends when compared to optical measurements and Ti II abundances have excessive scatter (\citealp{hawkins2016}, \citetalias{jonnson20}).

We further note the exclusion of some Na, K, Mn, and Ce abundances. As in previous data releases, Na remains one of the most imprecisely measured elements in DR16 with scatter of $\sim 0.1$ dex \citepalias{jonnson20}. As [Na/Fe] errors exceed 0.1 dex at low metallicities, we cut all stars with $\feh<-0.5$ in the Na analysis. While our quality cuts remove all stars below 3500 K, ASPCAP does not report calibrated abundances for Na, K, Mn, and Ce below $\sim$4000 K. Systematic trends with temperature in all four elements motivate their exclusion. Radial velocity shifts that move Ce lines into chip gaps or out of APOGEE's wavelength range cause additional omission. After our cuts the [Na/Mg]  high-Ia and low-Ia medians still have some temperature dependence. We find that lower temperature sub-samples have higher median [Na/Mg] values than higher temperature sub-samples at the same [Mg/H].  Appendix~\ref{ap:systematics} further explores this temperature systematic and its influence on the median trends of Na and other elements. 

\add{The median errors on the abundance [X/Mg] are comparable to those reported by \citet{jonnson20}. We find that the median errors are $\lesssim 0.06$ dex for all elements except Ce. Errors grow (to $\sim 0.1$ dex) in the lower metallicity regime for P, S, K, V, Cr, Co, and Cu. The reported abundance errors do not capture the spread in abundances about the median trends, suggesting some intrinsic abundance scatter (see Vincenzo et al., in prep for more details). The statistical errors on the median values themselves, though, are small, as we are taking medians of 100s of stars.}

\subsection{Systematics and Artifacts} \label{subsec:artifacts}

\begin{figure*}[!htb]
 \begin{centering}
 \includegraphics[width=\textwidth, angle=0]{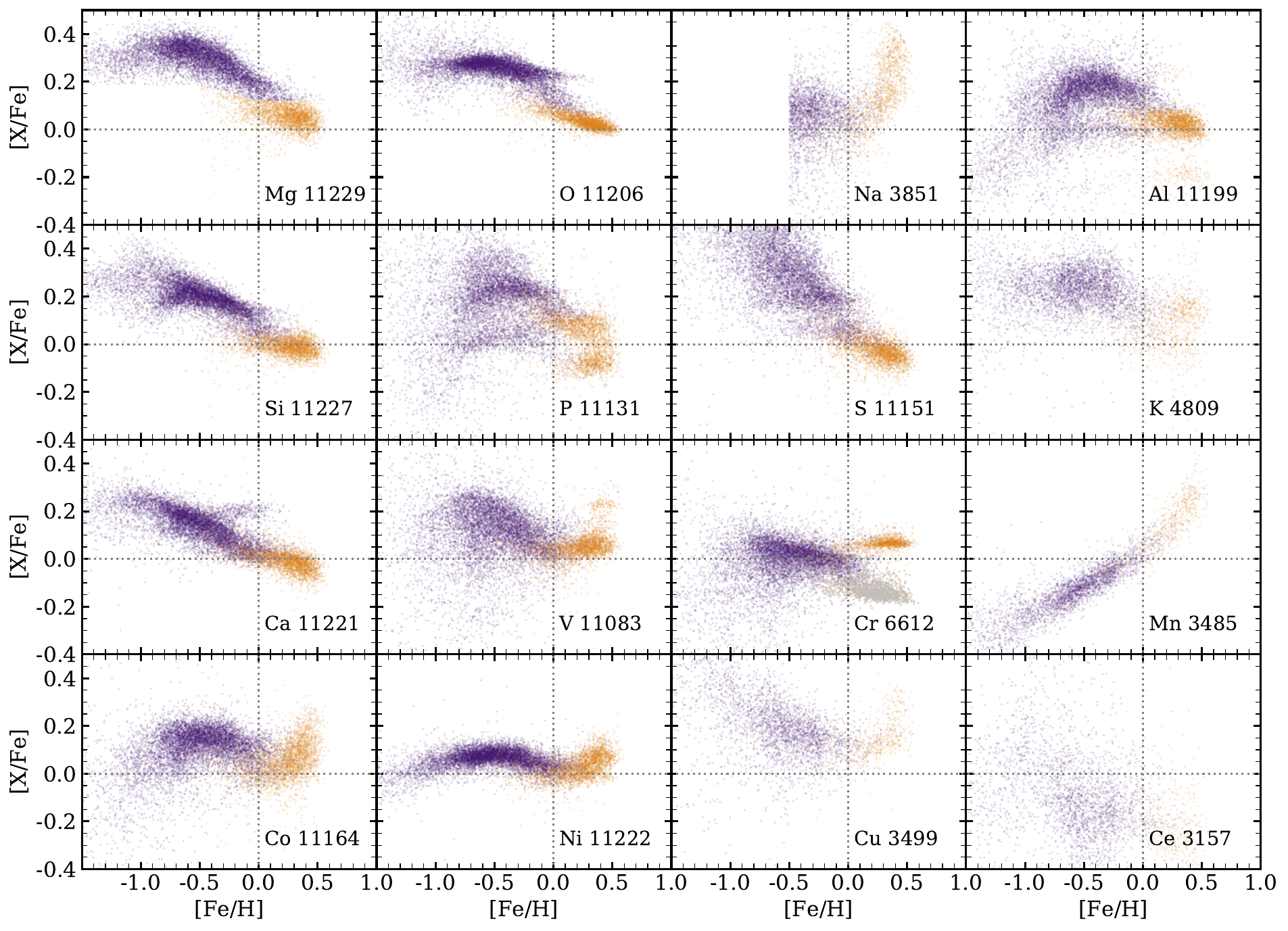}
 \caption{Bulge star distributions in [X/Fe] vs. [Mg/Fe] space for APOGEE elements. High-Ia stars are in orange and low-Ia stars are in purple. The dotted lines in each panel denote solar [X/Fe] and [Fe/H]. The number in the bottom right hand corner of each subplot corresponds to the number of stars for each elemental population which pass our cuts. Excluded Cr stars have been colored grey.}
 \label{fig:XFe}
 \end{centering}
\end{figure*}

Although we apply extensive quality cuts to our data, artifacts and unexpected abundance structures still appear. Most obviously, the high-Ia Cr stars split into two populations, one with higher [Cr/Mg] values, and one with lower. Clumps and bands can also be seen in [V/Mg], [P/Mg], and [Al/Mg]. Figure~\ref{fig:XFe} plots these same data, but in $\xfe$ vs. $\feh$ space. Here the aforementioned artifacts become more apparent and other structures not seen in $\xmg$ vs. $\mgh$ space pop out. We review the observed features that are caused by two broad problems in the data reduction process \citepalias{jonnson20} below. 

The $\alpha$ finger: In the [O/Fe], [Ca/Fe], and [Si/Fe] vs. $\feh$ panels of Figure~\ref{fig:XFe}, a `finger' can be seen protruding from the abundance tracks. \citet{zasowski2019} see a similar feature in their DR14 bulge abundance distributions. The finger feature affects a small fraction of cool giants and is caused by an error on the abundance determination \citepalias{jonnson20}.  It is also seen in the [$\alpha$/M] vs. [M/H] trends. Si exhibits a smaller finger than O and Ca, and Mg is free of the structure. While cutting stars below $3500$ K removed many of the finger stars from our analysis, the artifact still remains visible. However, due to the low number of stars in the finger, their exclusion changes the median high-Ia and low-Ia $\xmg$ trends by $<0.05$ dex (see Appendix~\ref{ap:systematics}). As such, we continue in our analysis with their inclusion. Interestingly, the finger artifact mainly plagues APOGEE bulge fields.

Cr bimodality: In both Mg and Fe space, the split in high-Ia Cr abundances is obvious. \citetalias{jonnson20} describe this artifact as the ``low abundance trend'' seen for some giant stars with $\teff < 4000$ K and suggest that it comes from the TIE-option in the FERRE interpolation. While this structure affects many elements, it afflicts Cr the most. The inclusion of both groups would create a high-Ia median trend that poorly describes either one. We choose to exclude the high-Ia stars with lower [Cr/Mg] from our analysis. These stars are located within the box drawn in the Cr panel of Figure~\ref{fig:XMg}, having $0<\mgh<0.75$ and $-0.3<$[Cr/Mg]$<-0.1$. We cut the low-Cr group because Cr trends in higher $\logg$ disk populations from APOGEE DR16 bear greater resemblance to the high-Cr track. Knowledge of previous works in the solar neighborhood \citep[e.g.][]{bensby2014} also support this decision. The excluded stars have been colored grey in Figure~\ref{fig:XFe}. They have $\femg$ near solar, so they shift location relative ot the low-Ia population, which has $\feh < 0$. 

Al and P banding: In both elemental plots, the high-Ia and low-Ia populations appear to be divided into multiple sequences or clumps. These are both instances of the low abundance trend artifacts, similar to that seen in Cr \citepalias{jonnson20}. We are unable to remove or account for these systematics as they are blended into the other features, so the abundance trends should be viewed with caution. We note that the stars within the bands are different for P, Al, and Cr.

V Clump: A small cluster of high-Ia stars can be seen at higher [V/Mg] and [V/Fe] than the majority of the population. These stars are not the same stars in the Al or P banding. As with the $\alpha$ finger, the few stars in this clump have little effect on the median trend, so we do not eliminate them. 

The inclusions/exclusions of stars afflicted by the artifacts discussed in this section slightly change the high-Ia and low-Ia median sequences with the only large impact being the high-Ia [Cr/Mg] trend. For a larger discussion of the effect that these stars have on the median trends, see Appendix~\ref{ap:systematics}. \citetalias{jonnson20} note that the low abundance trends can also be seen in Ni and Co, but this structure is not apparent in our stellar selection. 

\subsection{Bulge Definition and Bar Influence}\label{subsec:bar}

If the location and separation of the $\xmg$ median high-Ia and low-Ia trends differs between the bulge and the disk, implying some radial dependence, we might also expect the bulge definition to influence the observed median trends. In this paper we take the bulge to be all stars within the cylinder defined by $\rgc < 3 \kpc$ and $|Z|<5 \kpc$. Previous studies of the bulge and inner Galaxy have used similar definitions (see Section~\ref{sec:intro}), though not identical as the bulge/disk division is not clearly defined. We investigate the radial dependence of the abundance trends in the inner Galaxy by subdividing the bulge ($\rgc<3 \kpc$) into three groups—$2 \kpc \leq \rgc < 3 \kpc$, $1 \kpc \leq \rgc < 2 \kpc$, and $\rgc < 1 \kpc$. For each group, we calculate the median high-Ia and low-Ia trends and compare with the full sample. We find that the locations of the median trends are identical for all three groups within the uncertainties, though Mn, K, Cu, and Ce do not have enough stars on the high-Ia sequence to derive median trends for the innermost radial bin. We extend this comparison to the vertical bounds as well. Our bulge cut extends to high $|Z|$, so we divide the stars into two height bins of $|Z|<0.25 \kpc $ and $1 \kpc$ $ \leq |Z| < 5\kpc$ to probe the mid-plane and extended bulge. We again find that the high-Ia and low-Ia trends remain constant. We do not see a radial or height dependence of the high-Ia or low-Ia medians for stars within a radius of $3 \kpc$.

\begin{figure}[!htb]
\centering
 \includegraphics[width=\columnwidth]{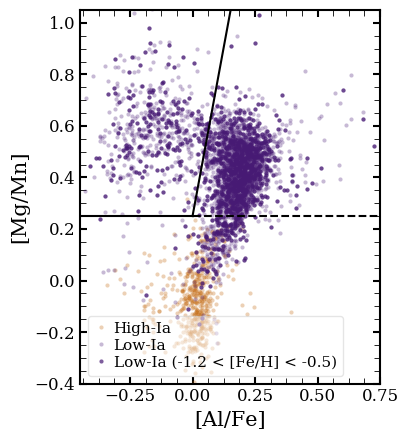}
 \caption{\add{[Mg/Mn] vs. [Al/Fe] distribution for $\sim 3400$ high-Ia (orange) and low-Ia (purple) stars. The solid and dashed lines are taken from \citet{horta2020}, where the solid lines separate stars with $-1.2<\feh<-0.5$ of accreted origin (high [Mg/Mn], low [Al/Fe]) from those of \textit{in situ} formation and the dashed line at [Mg/Mn] $= 0.25$ separates high and low-$\alpha$ stars. Our low-Ia stars with $-1.2<\feh<-0.5$ are plotted in dark purple.}}
 \label{fig:MnMg}
\end{figure}

\add{Our bulge definition is based on spatial location, and we do not attempt to distinguish between \textit{in situ} and accreted populations, or between bulge and co-located halo stars, or between heated disk stars ad stars born in a spheroidal configuration. Definitions that include kinematic selection or use abundance to isolate accreted populations would lead to a different bulge sample. However, the similarity of median trends at low $|Z|$ and higher $|Z|$, and the similarity of bulge and disk trends discussed below, suggest that alternative bulge definitions would not change our findings about median sequences, even though they might change the distribution of stars in [Mg/H] and [Fe/Mg]. \citet{das2020} and \citet{horta2020} use the [Mg/Mn] vs. [Al/Fe] diagram to identify stars with $-1.2<\feh<-0.5$ that they suggest represent an accreted population. Figure~\ref{fig:MnMg} shows this diagram for the $\sim 3400$ stars of all metalicities in our sample that have the most reliable Mn abundances. We find $\sim 450$ stars in the region of parameter space that these papers describe as ``the blob'' and ``the Inner Galaxy Structure,'' respectively. About 140 of these stars fall into the $-1.2<\feh<-0.5$ metalicity range and thus may have been accreted. We refer to the above papers for further investigation of the kinematic and chemical origins of these stars. }

Additionally, the Milky Way bar may have a distinct chemical signature. Work by \citet{bovy2019} found that inner galaxy stars ($\rgc<5\kpc$) inside and outside of the bar followed the same [O/Fe] vs. $\feh$ tracks, but that there were more metal poor stars ($\feh <0$) in the bar than the outside of the bar. 
Lian et al. (in prep) find similarly small differences between the [Fe/H] and [Mg/Fe] distributions on and off the inner bar. We investigate whether the $\xmg$ vs. $\mgh$ trends differ between on and off bar inner Galaxy regions. We use the bar definition from \citet{bovy2019}: an ellipse with a semi major axis of 5 kpc angled at 25$^{\circ}$ clockwise from the Sun-Galaxy line with an axis ratio of 0.4. We take the ~2000 stars in this ellipse with $\rgc<3 \kpc$ and find the median high-Ia and low-Ia abundance trends for each element with sufficient observations. While the metallicity distribution varies, we find good agreement between the median tracks of the bar and the full inner galaxy population. Small differences (~0.05 dex) are seen at the low $\mgh$ end of the low-Ia sequence for Al, Si, V, Cr, and Co. Many of these elements show the low abundance artifacts noted by \citetalias{jonnson20}, so the deviations are likely due to the small number of stars in the sample and the influence of these artifacts.

\subsection{Creating a Comparison Sample}\label{subsec:sample}

\begin{figure*}[!htb]
\centering
 \includegraphics[width=.8\textwidth]{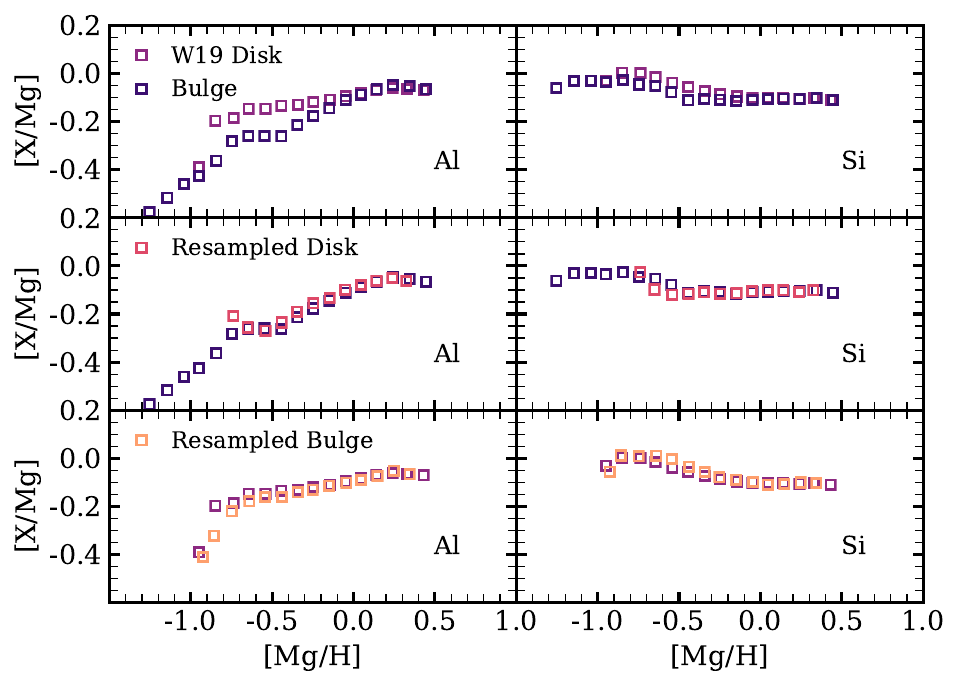}
 \caption{Median low-Ia trends for Al and Si, comparing different stellar samples. Top row: Bulge medians (dark purple) compared to W19 disk medians (light purple). Middle row: Bulge medians (purple) compared to disk stars of $5 \leq R_G \leq 11$ whose $\log(g)$ distribution matches that of the bulge (resampled disk; pink). Bottom row: W19 disk medians (pink) compared to a subset of stars in the bulge sample that reproduce the $1 < \log(g) < 2$ distribution of the W19 disk (resampled bulge; orange). After $\log(g)$ systematics are accounted for in the lower two rows, the differences seen in the top row disappear.}
 \label{fig:AlSiMeds}
\end{figure*}

By comparing the median abundance trends of the bulge and disk, we can learn about the similarities and/or differences in their chemical enrichment histories. Upon first analysis of the bulge and W19 disk medians, we see differences of $\sim 0.1-0.2$ dex in half of the elements. The first row of Figure~\ref{fig:AlSiMeds} shows those of Al and Si. The low-Ia [Al/Mg] bulge sequence dips below that of the W19 disk at sub-solar $\mgh$. The differences in Si are subtler, but both the bulge high-Ia and low-Ia medians appear suppressed relative to the disk.

\begin{figure*}[!htb]
 \includegraphics[width=\textwidth]{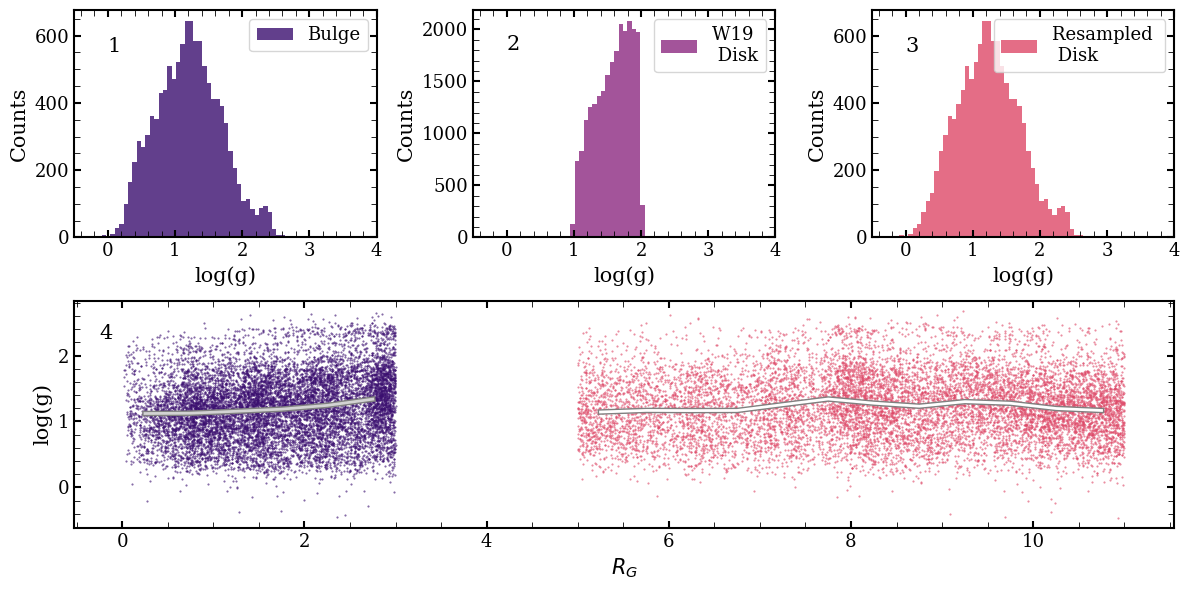}
 \caption{Top row: $\log(g)$ distributions for stars in the bulge (left; dark purple), the W19 disk sample (middle; purple), and the resampled disk (right; pink). Bottom: Radial location of stars with a given $\log(g)$ for the bulge and the resampled disk populations. We overplot median trends binned radially by 0.5 kpc. Bulge and disk stars appear evenly radially distributed with no clear dependence on $\log(g)$.}
 \label{fig:bulge_dems}
\end{figure*}

These two populations, however, do not probe similar stellar samples. APOGEE observes different stellar samples at different locations in the Galaxy. As the bulge is farther away, APOGEE observes only the most luminous giants. This means that the $\logg$ distribution of stars differs between the bulge and the W19 disk (Figure~\ref{fig:bulge_dems}, panels 1 and 2). Systematic abundance errors that correlate with $\logg$ could cause artificial differences between the bulge and disk medians \citep[e.g.][]{santos-peral2020}. To test this possibility, we randomly resample the full disk population to reflect the $\logg$ distribution of the bulge. To ensure that we sample disk stars and not inner galaxy stars on the bulge border, we restrict our disk to $5 \kpc < \rgc < 11 \kpc$. In each $\logg$ bin, we select a number of APOGEE disk stars equal to the number of bulge stars in the same bin. We will hereafter refer to this stellar population as the ``resampled disk''. Figure~\ref{fig:bulge_dems} plots histograms of the bulge and resampled disk's $\logg$ distribution in panels 1 and 3 of the top row and their radial position in the bottom row (panel 4). Stars in both populations are distributed throughout the bulge and disk, respectively.

After this re-sampling, the median $\xmg$ vs. $\mgh$ high-Ia and low-Ia trends come into closer agreement for many elements. The middle panel of Figure~\ref{fig:AlSiMeds} compares [Al/Mg] and [Si/Mg] vs. $\mgh$ medians for the bulge and the $\logg$ resampled disk. Differences seen between the W19 disk and bulge almost disappear. To confirm that this is a result of $\logg$ systematics and dissimilar sampling we repeat for the opposite case--randomly resampling the bulge to reflect the $\logg$ distribution of the W19 disk. We again see that the median [Al/Mg] and [Si/Mg] vs. $\mgh$ trends agree between the two population with similar $\logg$ distributions (Figure~\ref{fig:AlSiMeds}, bottom row). We conclude that $\logg$ systematics cause small but detectable changes in the elemental median trends. In order to accurately compare Galactic regions, we must have samples with similar $\logg$ distributions. Our bulge and resampled disk populations satisfy this requirement. 

\add{This resampling also removes the bias from NLTE corrections. \citet{jonnson20} note that elements such as Mg, Al, K, and Ca might be influenced by NLTE effects not accounted for in APOGEE's current data reduction pipeline. These effects may be responsible for shifts in abundances between dwarf and giant stars. Our $\logg$ constraints ensure that our bulge and disk populations have the same stellar makeup, so a relative abundance comparison should see no influence from NLTE effects.}

\begin{figure}[!htb]
 \includegraphics[width=\columnwidth]{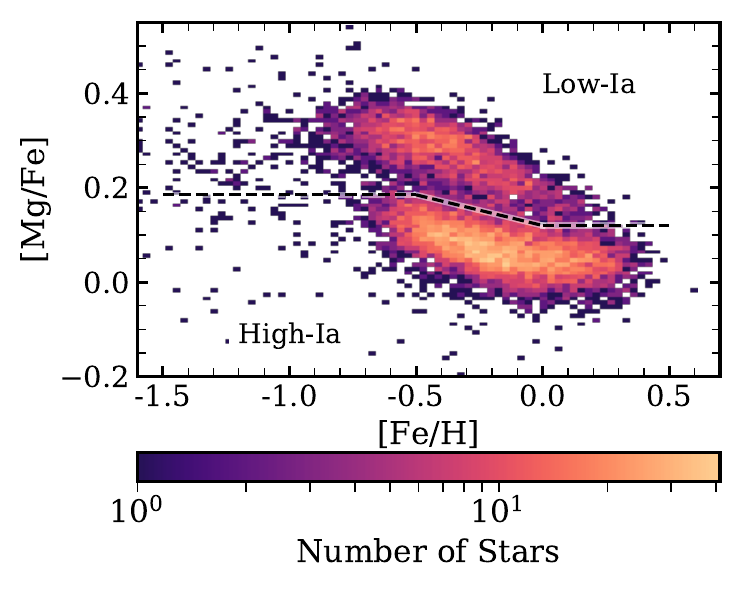}
 \caption{Density plot of the stars in the resampled disk that recreate the $\log(g)$ distribution of the bulge. The dashed line denotes our division between the high-Ia and low-Ia stars used in the bulge. The two sequences can clearly be distinguished, as in Figure 1 of W19.}
 \label{fig:disk_den}
\end{figure}

A density plot of our resampled disk can be found in Figure~\ref{fig:disk_den}. We use the same low-Ia definition as in Equation (\ref{eq:boundary}). The high-Ia and low-Ia populations are clearly defined and separated. The resampled disk has more stars on the high-Ia sequence and a less extended low-Ia sequence than the bulge, reflective of the distribution found by \citetalias{weinberg}. We apply the same exclusions to the resampled disk as described above.

\subsection{Comparison to APOGEE Disk} \label{subsec:comparison}

\begin{figure*}[]
 \begin{centering}
 \includegraphics[width=1.2\textwidth, angle=90]{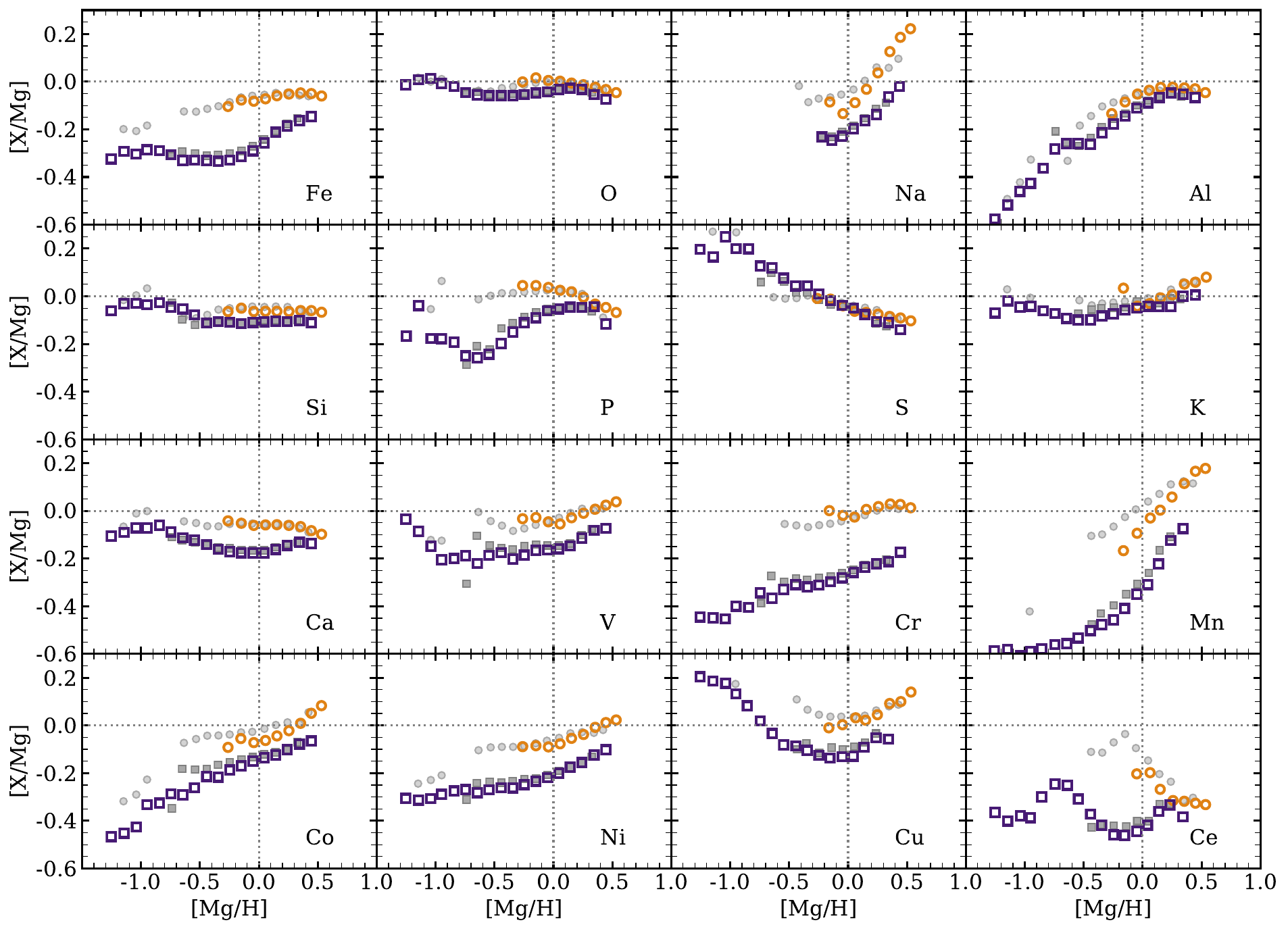}
 \caption{APOGEE bulge star median abundances trends of high-Ia (orange circles) and low-Ia (purple squares) populations, as in Figure 2. Median high-Ia and low-Ia abundances of resampled disk stars are plotted as light grey circles and dark grey squares, respectively, for comparison. Dotted lines correspond to solar values of [X/Mg] and [Mg/H].}
 \label{fig:comp_bulge_disk}
 \end{centering}
\end{figure*}

Now that we have similar stellar samples, we can compare the high-Ia and low-Ia median abundance trends of the bulge and the resampled disk. Figure~\ref{fig:comp_bulge_disk} plots the median bulge trends (same as Figure~\ref{fig:XMg}) and the median trends of the resampled disk. Both are binned by 0.1 dex in $\mgh$, requiring $>20$ stars per bin. Overall, the medians agree well. The resampled disk and bulge trends are nearly identical for Fe, O, Al, Si, S, K, Ca, V, Cr, Ni, and Cu. This similarity suggests that the bulge and the disk experienced similar chemical enrichment and that nucleosynthesis pathways are identical throughout the Galaxy--extending the conclusions of \citetalias{weinberg} from the disk to the inner Galaxy. We do see minor differences between the Na, P, Mn, Co, and Ce trends of bulge and resampled disk. All of these difference, though, are on the scale of 0.1 dex or smaller (see Figure~\ref{fig:medDiffsIMF}). In the following subsections we will more closely examine the median trends, discussing APOGEE's elemental precision, nucleosynthetic origins, and differences/similarities between the bulge and the resampled disk. We will reference the chemical evolution models of \citet{andrews17}, hereafter AWSJ17, for theoretical yield predictions. 

\subsubsection{$\alpha$-elements}
CCSN dominate the production of $\alpha$-elements such as O, Si, S, and Ca. O is predicted to come almost entirely from CCSN, as is Mg, while Si, S, and Ca have some SNIa contribution \citepalias{andrews17}. As such, we see that O traces Mg and has little to no separation between the high-Ia and low-Ia [O/Mg] medians. APOGEE's  [O/Mg] trends are flat, supporting metallicity independent CCSN yields. This agrees with other IR studies \citepalias{jonnson20} but is inconsistent with metallicity dependent [O/Mg] trends derived from optical abundances \citepalias{griffith19}. 

Si and Ca show larger median high-Ia and low-Ia separations than O, in agreement with theoretical predictions that they have a non-dominant SNIa contribution. We note that Si is one of the most precisely measured elements in APOGEE with scatter of $\lesssim 0.02$ dex \citepalias{jonnson20}. While \citetalias{andrews17} find that S has a larger SNIa contribution than Si, we see no separation between the S high-Ia and low-Ia median trends, indicative of little SNIa contribution. S is the only $\alpha$-element for which we infer metallicity dependent yields.

The high-Ia and low-Ia median trends of the $\alpha$-elements agree very well between the bulge and the resampled disk. The median absolute differences between the sequences are all $\leq 0.02$ dex and within observational errors. We do not see the enhancement in the bulge $\alpha$ abundances relative to the local thick disk at a give metallicity, as was found in previous works \citep[e.g.][]{bensby2013, johnson2014}

\subsubsection{Light odd-$Z$ elements}

Like the $\alpha$-elements, we expect CCSN to be the predominant source of the light odd-$Z$ elements (Na, Al, P, and K) and expect SNIa to contribute in a non-dominant way. The abundance tracks of these elements, however, should be distinguished from $\alpha$-elements by stronger metallicity dependence \citepalias{andrews17}. While [Na/Mg] high-Ia and low-Ia median trends do show a positive metallicity gradient, they also show a significant separation. This suggests a large SNIa or other delayed production source. The median trend separation agrees with results for the APOGEE DR14 disk \citepalias{weinberg} and GALAH DR2 sample \citepalias{griffith19}. Both papers include a larger discussion of the nucleosynthetic implications of the empirical results. We see a relatively similar sequence separation in the Na trends of the resampled disk and the bulge, but the two samples have different high-Ia metallicity dependencies. While the low-Ia tracks are in good agreement, the high-Ia trends have a median absolute difference of 0.06 dex, one of the largest differences between the bulge and the resampled disk. We reiterate that Na is not precisely determined by APOGEE (with scatter of $\sim 0.1$ dex), and that temperature systematics have not been accounted for \citepalias{jonnson20}. The difference between bulge and disk trends for the high-Ia population could be a result of different chemical enrichment histories, but it could also be a result of residual systematic errors on Na abundance measurements. 

Al and P are also plagued by systematics and reduction artifacts. Both show some separation in their median trends, though not to the extent of Na. This separation is unexpected for Al, which is predicted to be a pure CCSN element \citepalias{andrews17}. \citetalias{weinberg} find no separation in their analysis of DR14 data. Al trends do show a positive metallicity dependence, as expected. The bulge and resampled disk [Al/Mg] trends agree well after the disk resampling. Al abundances have high precision ($\lesssim 0.04$ dex) in APOGEE DR16 \citepalias{jonnson20}.

The [P/Mg] trends show a 0.05-0.1 dex differences between the bulge and the resampled disk in some $\mgh$ bins.  A closer examination of stars in the high-Ia and low $\mgh$ bins--most discrepant between the bulge and the disk--reveals that the $\logg$ distribution of stars in the respective populations differ significantly. While the resampled disk stars span the parent sample's $\logg$ range of 0-2.5, the less populated bulge bin only includes stars near $\logg = 0$, perhaps because only the most luminous stars yield successful P measurements at low $\mgh$ at these distances. The $\logg$ systematics could be responsible for some of the median bin differences. We reiterate that P is the least precisely determined element in APOGEE (with scatter of 0.15 dex or greater) and is subject to systematic errors \citepalias{jonnson20}. \citet{masseron2020}, for example, find that low-metallicity P-rich stars are missed by ASPCAP entirely. We do not trust the P abundances and do not draw conclusions from these median trends.

We see the weakest metallicity dependence of the light odd-$Z$ elements in the [K/Mg] median trends. The high-Ia and low-Ia medians show little separation and suggest CCSN dominated production. This agrees with theoretical yields \citepalias{andrews17}. We see good agreement between the bulge and the resampled disk, though the resampled disk's low-Ia median trend sits at a higher [K/Mg] than the bulge low-Ia median.

\subsubsection{Fe-peak elements}

Nucleosynthetic yields and chemical evolution models predict comparable CCSN and SNIa contribution to Fe-peak elements at solar abundances \citepalias{andrews17}. Our observations agree, as we see significant separation between the high-Ia and low-Ia median trends for all APOGEE Fe-peak elements. \citetalias{andrews17} further predict that the odd-$Z$ elements (V, Mn, Co) will have a positive metallicity dependence while the even-$Z$ elements (Cr, Fe, Ni) will have flatter trends. We observe Mn and Co to have a stronger metallicity dependence than the even-$Z$ elements. V exhibits a shallower trend than Co and Mn. [V/Mg] and [Co/Mg] have similar, small separation between the high-Ia and low-Ia median trends, indicative of significant but non-dominant SNIa contribution. As in \citetalias{weinberg} and \citetalias{griffith19}, Mn has the largest separation of all included elements, and thus the largest SNIa contribution. Among the even-$Z$ elements, [Cr/Mg] trends show larger separation than [Fe/Mg] (with the caveat that we cut out low-[Cr/Mg], high-Ia stars as discussed in Section~\ref{subsec:artifacts}), while [Ni/Mg] trends show less separation. This suggests that SNIa dominate Cr production, while CCSN make a larger contribution to Ni. Both elements strongly resemble Fe.

The V, Cr, Fe, and Ni median trends agree well between the bulge and resampled disk. All have absolute median differences $\leq 0.02$ dex. Both populations' $\femg$ low-Ia median sequences plateau around $-0.3$. Mn and Co show larger differences. Both the high-Ia and low-Ia resampled disk [Mn/Mg] median trends sit at slightly higher [Mn/Mg] than those of the bulge and have a flatter metallicity dependence. The bulge and resampled disk's high-Ia and low-Ia [Mn/Mg] trends have median absolute differences of 0.07 and 0.05 dex respectively. We reiterate that APOGEE DR16 Mn abundances are not populated for cool stars ($T \lesssim 4000$ K) due to temperature systematics, so we have a much smaller sample than the rest of the bulge with few high-Ia stars. While cluster calibrations were applied to DR14 data to remove temperature effects \citep{holtzman18}, these calibrations were not done in DR16. Mn shows some of the largest differences between the bulge and the resampled disk, particularly for high-Ia stars, but due to the small bulge sample and potential temperature systematics we interpret these results cautiously. The Co high-Ia and low-Ia resampled disk medians are slightly higher in [Co/Mg], with flatter metallicity dependence than the bulge. The deviations in the Co low-Ia trends seen at sub-solar $\mgh$ can also be seen, to a much smaller degree, in Ni, and V. 

\subsubsection{Cu and Ce}

In \citetalias{griffith19}, we classify Cu as a ``Fe-cliff'' element, as it resides on the steeply falling edge of the Fe abundance peak. APOGEE DR16+ Cu abundances are not the most reliable, having a low precision and low accuracy \citepalias{jonnson20}. The upturned tail at low $\mgh$ and $\feh$ seen in Figures~\ref{fig:XMg} and~\ref{fig:XFe} is unexpected and probably not trustworthy. We will limit our interpretation to Cu abundances above solar $\mgh$. Here we see separation between the high-Ia and low-Ia sequences and a positive metallicity dependence. Cu production is thought to be dominated by CCSN, with yields suggesting a strong metallicity dependence \citepalias{weinberg}. The observed separation in median trends, however, implies a non-zero SNIa or other delayed component, in agreement with optical trends from GALAH \citepalias{griffith19}. At all metallicities, we see good agreement between the bulge and resampled disk median trends. 

While all of the above elements are likely produced through CCSN and SNIa nucleosynthesis, Ce is produced by a combination of the rapid and slow neutron capture processes ($r$-process and $s$-process, respectively) \footnote{The $r$-process and $s$-process do not automatically translate into prompt and delayed, since, for example, some $s$-process production occurs in massive CCSN progenitors. In practice, most $s$-process enrichment arises in AGB stars with intermediate lifetimes, and the main sources of $r$-process enrichment appear to be prompt.}. Separation in the Ce high-Ia and low-Ia median trends indicates that Ce is dominated by a delayed enrichment source, in agreement with measurements by \citet{arlandini99} and \citet{bisterzo14} who find Ce to have $\sim80\%$ $s$-process contribution. We see similar sequence separation between the bulge and the resampled disk, though the resampled disk's low-Ia trend has slightly higher [Ce/Mg] than that of the bulge. This difference should be viewed with some caution, as APOGEE Ce abundances have low precision (scatter of 0.15 dex or greater) due to its single line analysis \citepalias{jonnson20}. We do, however, think that Ce abundances are should be accurate (excluding low metallicity stars). Neutron capture elements are not well studied in the bulge, so APOGEE Ce observations can give insight into nucleosyntheic processes beyond CCSN and SNIa. 

DR16 is the first APOGEE data release to include Cu and Ce. A larger discussion of their abundance trends, nucleosynthetic implications, and comparison to GALAH data can be found in Appendix~\ref{ap:Ce_Cu}. 

\section{Two-process Model}\label{sec:two-proc}

With the exception of Ce, the elements studied in this paper are theoretically expected to originate almost entirely from CCSN or SNIa \citep{johnson2019}. While individual CCSN produce elements in different ratios depending on progenitor mass, the IMF-averaged production should yield fixed ratios for supernovae of a given metallicity. \citetalias{weinberg} interpreted the median abundance-ratio sequences from APOGEE DR14 in terms of a two-process model, which represents the elemental abundances of any given star (or median sequence point) as the sum of a CCSN process with amplitude $\Acc$ and a SNIa process with amplitude $\AIa$: 
\begin{equation} \label{eq:XH}
    \bigg(\frac{\rm X}{\rm H}\bigg) = \Acc \pcc(Z) + \AIa \pIa(Z).
\end{equation}
To account for metallicity-dependent yields, the process vectors $\pcc$ and $\pIa$ are allowed to depend on the star's metallicity. \citetalias{griffith19} extend the analysis to GALAH DR2 abundance ratios, including elements not measured by APOGEE. In \citetalias{weinberg}, \citetalias{griffith19}, and here, the metallicity dependencies are modeled as power laws in (Mg/H), with index $\acc$ and $\aIa$ for $\pcc$ and $\pIa$, respectively. 

Here we fit the DR16 abundance-ratio sequences of the resampled disk to infer the two-process model parameters $\acc$, $\aIa$, and 
\begin{equation} \label{eq:RIa}
    \RIa = \frac{\pIasun}{\pccsun} \, ,
\end{equation}
which is the ratio of the two processes for element X in a star with solar abundances $\mgh = \femg = 0$. Differences from W19 arise partly from the differences between DR14 and DR16 abundance determinations and partly from the different $\logg$ distributions of the samples. The values of $\RIa$, $\acc$, and $\aIa$ derived here are not necessarily more reliable than those of \citetalias{weinberg} but they allow us to predict bulge stars abundances using a model ``trained'' independently on the disk. In Section~\ref{sec:IMFs}, we use these comparisons to place rough limits on the possible difference between the high mass IMF slope of bulge and disk populations.

As in \citetalias{weinberg}, we assume that Mg is purely produced by CCSN with metallicity independent yields ($R_{\text{Ia}}^{\text{Mg}}=0$ and $\alpha_{\text{cc}}^{\text{Mg}}=\alpha_{\text{Ia}}^{\text{Mg}}=0$), that Fe is produced by both SNIa and CCSN with metallicity independent yields ($\alpha_{\text{cc}}^{\text{Fe}}=\alpha_{\text{Ia}}^{\text{Fe}}=0$), and that stars on the low-Ia plateau, at $\mgfe \approx 0.3$, have pure CCSN enrichment (implying $R_{\text{Ia}}^{\text{Fe}}=1$). 
Therefore, given any star's $\mgfe$, the ratio of its SNIa to CCSN enrichment is
\begin{equation} \label{eq:aratio}
	\frac{\AIa}{\Acc} = 10^{0.3 - \mgfe} -1 \, .
\end{equation}
From the global elemental parameters ($\RIa$, $\acc$, $\aIa$) and stellar abundances ($\mgh$ $\mgfe$), \citetalias{weinberg} derive an expression to calculate the expected abundance of element $X$ in a star as
\begin{multline} \label{eq:xmg}
	\xmg = \acc \mgh + \\
	\log \bigg[ \frac{1+\RIa (\AIa/\Acc)  10^{(\aIa-\acc)\mgh} }{1 + \RIa} \bigg], 
\end{multline}
with $\AIa/\Acc$ from Equation (\ref{eq:aratio}). As in \citetalias{weinberg} and \citetalias{griffith19}, we will use the median $\xmg$ and $\mgh$ trends to find the best $\RIa$, $\acc$, and $\aIa$ values for each element. For elements that have substantial production by AGB stars (Ce and possibly Na, P, Cu), the model parameters are only qualitatively meaningful since the AGB enrichment delay is different from that of SNIa. 

\subsection{Fitting the Resampled Disk}\label{subsec:fitting_2proc}

We perform an unweighted, least-squares fit to the resampled disk's high-Ia and low-Ia $\mgh$ vs. $\xmg$ median trends for each element with the two-process model to derive the best fit $\RIa$, $\acc$, and $\aIa$ values. We allow all three parameters to vary and conduct a grid search with a step size of 0.01 for each. Due to the low number of stars at low metallicity, we only fit median points with $\mgh > -0.7$. We choose to fit the resampled disk rather than bulge because the resampled disk medians track the bulge medians and its high-Ia sequence is more well populated. 

As pointed out by \citetalias{griffith19} (see their Figure 10), the two-process model necessarily predicts [X/Mg] = 0 for a star, or median sequence point, with $\femg = \mgh = 0$. Figure~\ref{fig:comp_bulge_disk} shows that this is not the case for all elements in our sample, but the small departures from solar $\xmg$ on the high-Ia sequence are plausibly a consequence of small calibration errors in the abundance determinations. Following \citetalias{griffith19}, therefore we apply zero-point offsets to all elements in order to force the high-Ia sequence to pass through $\xmg=0$ at $\mgh=0$ prior to performing the fits. These offsets are listed in Table~\ref{tab:zeros} and are applied to the bulge and resampled disk sequences in all further analysis. We note that APOGEE does already apply zero point offsets to force stars with solar [M/H] in the solar neighborhood to have a mean [X/M]=0 \citepalias{jonnson20}, but the $\logg$ distribution of that sample is different from ours. (See \citetalias{griffith19} Figure 10 for an example of two-process fits with and without zero-point offsets.)

Under the two process model, elements solely produced by CCSN should have no separation between their high-Ia and low-Ia median sequences and should be best fit by $\RIa \approx 0$. As the SNIa contribution to an element increases, the sequence separation should also increase, driving the $\RIa$ value up. Elements with $\RIa \approx 1$ are produced equally by CCSN and SNIa in stars of solar abundance, and SNIa dominate the production of those with $\RIa > 1$. The $\RIa$ value can be converted to a CCSN fraction by the equation
\begin{equation}\label{eq:fcc}
    \fcc = \frac{1}{(1+\RIa)} \, .
\end{equation}
Elements with $\acc \approx \aIa \approx 0$ and $\RIa \approx 1$ will have median $\xmg$ trends that follow the corresponding $\femg$ sequences. As $\acc$ increases, the slope of both sequences should also increase, with positive/negative $\acc$ representing a positive/negative metallicity dependence. The $\aIa$ parameter will further change the slope of the high-Ia sequence relative to the low-Ia sequence.

\begin{figure*}[!htb]
\begin{centering}
\includegraphics[width=\textwidth, angle=0]{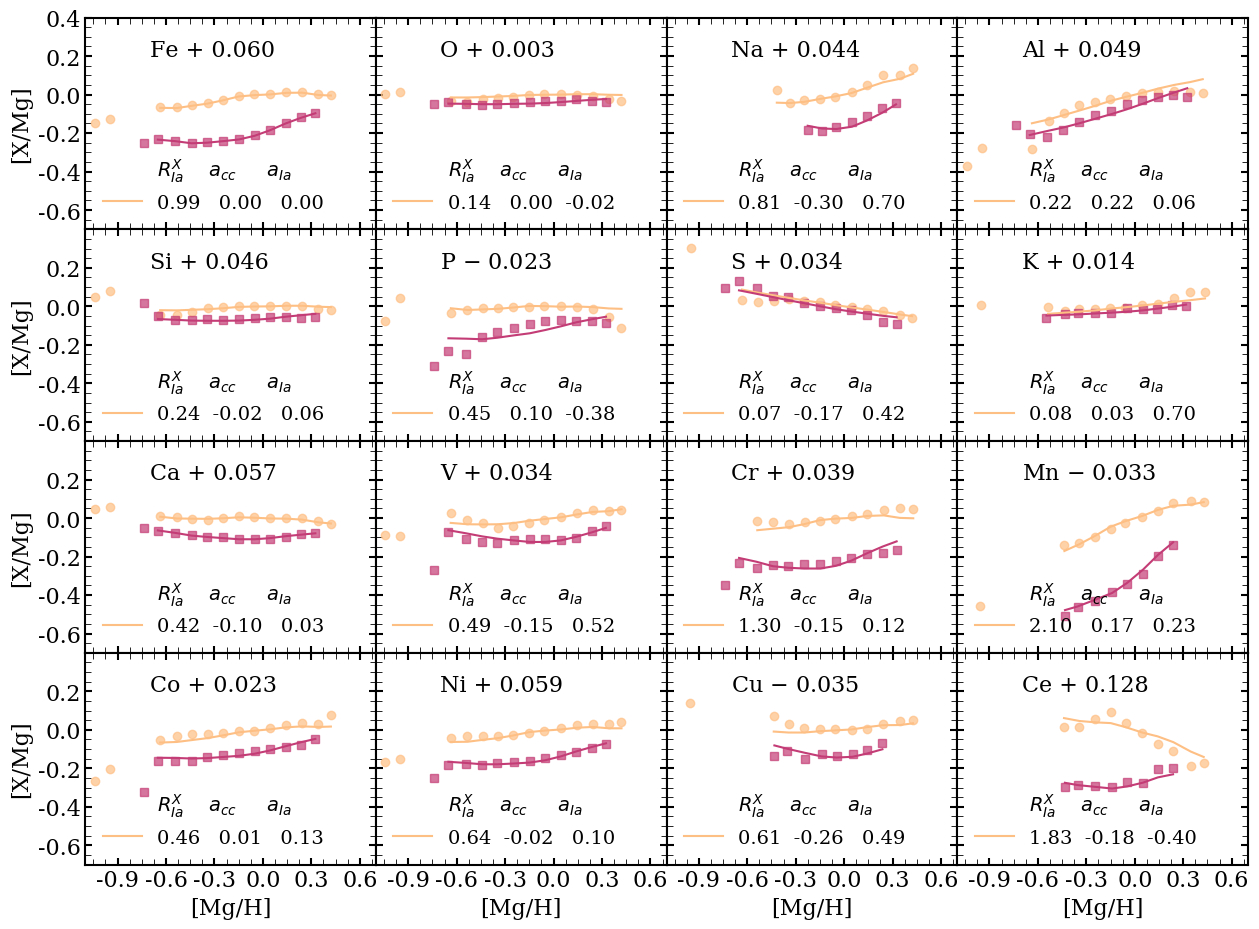}
\caption{Two-process model fits to the resampled disk median trends. Orange/pink circles/squares represent the high/low-Ia population medians including offsets. Solid lines are the best fit two-process sequences. Best fit sequences only extend to a lower bound of $\mgh=-0.7$ as fits were computed on median points with $\mgh\geq -0.7$. Elemental best fit two-process values and zero-point offsets are given in each cell.}
\label{fig:2proc}
\end{centering}
\end{figure*}

Figure~\ref{fig:2proc} shows the two-process model fits to the resampled disk's high-Ia and low-Ia medians (including an offset). The $\RIa$, $\acc$, and $\aIa$ values for each element as well as the zero-point offset applied prior to the fit are listed in the respective panels and in Table~\ref{tab:zeros}. The two-process model well describes the high-Ia and low-Ia trends of the resampled disk down to $\mgh = -0.7$ for most elements. However, we see that the Al, P, and Cr high-Ia medians are poorly fit at the high $\mgh$ end and that the two-process model is unable to reproduce the non-linear metallicity dependence of the [Cu/Mg] and [Ce/Mg] trends. If we allowed arbitrary metallicity dependencies, then the two-process model would be able to fit the observed median sequences by construction, so imperfect fits are a consequence of the power-law restriction. Comparisons like Figure~\ref{fig:2proc} are not in themselves a strong test of the two-process model's underlying assumptions; rather, these fits allow us to convert the observed sequences into physically meaningful quantities given those assumptions. Better tests of the model's validity come from predicting the abundance trends of other stellar populations or from predicting star-by-star deviations from median trends, which we will examine briefly in Section~\ref{subsec:predicting} and in more detail in future work.

\begin{deluxetable}{lrrrrr}[]
\tablecaption{Zero-point offsets and the best fit two-process model parameters for each element as fit to the median high-Ia and low-Ia points of the resampled disk. $\fcc$ denotes the fractional CCSN contribution, as defined in Equation (\ref{eq:fcc}).\label{tab:zeros}}
\tablehead{
\colhead{[X/Mg]} & \colhead{Offset} & \colhead{$\RIa$} & \colhead{$\acc$} & \colhead{$\aIa$} & \colhead{$\fcc$}
}
\startdata
Fe 	 & $ 0.060 $ & $ 0.99 $ & $  0.00 $ & $  0.00 $ & $ 0.503 $ \\ 
O 	 & $ 0.003 $ & $ 0.14 $ & $  0.00 $ & $ -0.02 $ & $ 0.877 $ \\ 
Na 	 & $ 0.044 $ & $ 0.81 $ & $ -0.30 $ & $  0.70 $ & $ 0.552 $ \\ 
Al 	 & $ 0.049 $ & $ 0.22 $ & $  0.22 $ & $  0.06 $ & $ 0.820 $ \\ 
Si 	 & $ 0.046 $ & $ 0.24 $ & $ -0.02 $ & $  0.06 $ & $ 0.806 $ \\ 
P 	 & $ -0.023 $ & $ 0.45 $ & $  0.10 $ & $ -0.38 $ & $ 0.690 $ \\ 
S 	 & $ 0.034 $ & $ 0.07 $ & $ -0.17 $ & $  0.42 $ & $ 0.935 $ \\ 
K 	 & $ 0.014 $ & $ 0.08 $ & $  0.03 $ & $  0.70 $ & $ 0.926 $ \\ 
Ca 	 & $ 0.057 $ & $ 0.42 $ & $ -0.10 $ & $  0.03 $ & $ 0.704 $ \\ 
V 	 & $ 0.034 $ & $ 0.49 $ & $ -0.15 $ & $  0.52 $ & $ 0.671 $ \\ 
Cr 	 & $ 0.039 $ & $ 1.30 $ & $ -0.15 $ & $  0.12 $ & $ 0.435 $ \\ 
Mn 	 & $ -0.033 $ & $ 2.10 $ & $  0.17 $ & $  0.23 $ & $ 0.323 $ \\ 
Co 	 & $ 0.023 $ & $ 0.46 $ & $  0.01 $ & $  0.13 $ & $ 0.685 $ \\ 
Ni 	 & $ 0.059 $ & $ 0.64 $ & $ -0.02 $ & $  0.10 $ & $ 0.610 $ \\ 
Cu 	 & $ -0.035 $ & $ 0.61 $ & $ -0.26 $ & $  0.49 $ & $ 0.621 $ \\ 
Ce 	 & $ 0.128 $ & $ 1.83 $ & $ -0.18 $ & $ -0.40 $ & $ 0.353 $ \\
\enddata
\end{deluxetable}

\citetalias{weinberg} report two-process fits to the APOGEE DR14 disk for all elements included here except Cu and Ce (new to DR16). They discuss each element, its two-process model fit and the nucleosynthetic implications, and they compare the derived $\fcc$ value to theoretical results from \citet{rybizki17}. We find very similar $\fcc$ values (a difference of $\leq 10\%$) for Fe (by construction), and for O, Na, Si, S, Ca, Cr, Mn, and Ni, so we do not go into detail about the two-process model implications in this paper. We discuss the two-process model implications for Cu and Ce in Appendix~\ref{ap:Ce_Cu}.

\begin{figure*}[htb]
 \begin{centering}
 \includegraphics[width=1\textwidth, angle=0]{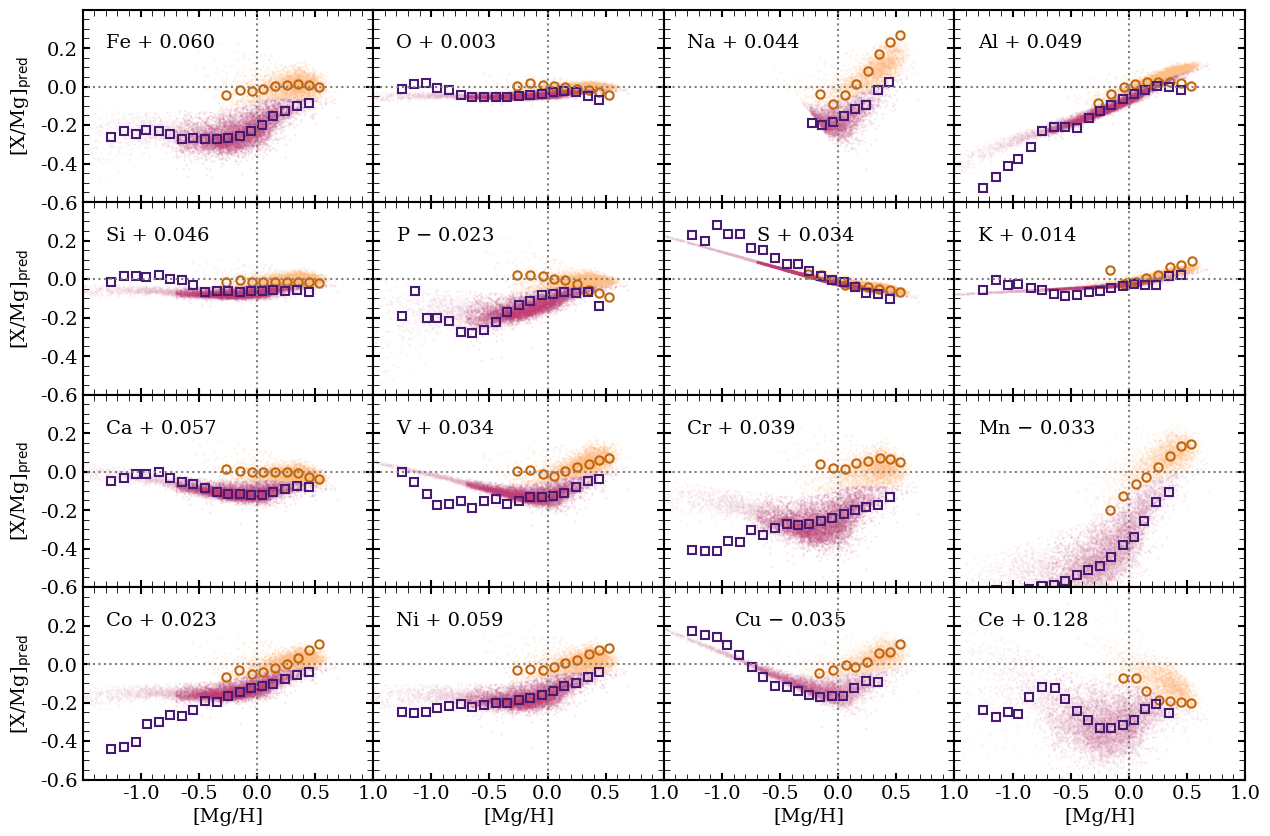}
 \caption{Predicted abundances for the bulge population given the resampled disk's two-process model parameters. For each bulge star, $\xmg$ values are predicted based on the star's measured $\mgh$ and $\femg$. Low-Ia/high-Ia stars are colored pink/orange respectively. Lighter shades indicate extrapolated abundance predictions ($\mgh<-0.7$). The bulge median observed high-Ia and low-Ia trends, including offsets listed in the top left corner, are over-plotted in dark orange circles and purple squares, respectively.}
 \label{fig:pred_meds}
 \end{centering}
\end{figure*}

We do, however, see differences between our derived two-process fits and those of \citetalias{weinberg} for Al, P, K, V, and Co . In all of these cases, we have repeated our fits using the DR16 abundances but the disk $\logg$ and geometry cuts used by \citetalias{weinberg}. The differences persist, indicating that they arise primarily from changes in the ASPCAP abundance pipeline, not from differences in the $\logg$ distribution or location of the sample. For Al, the higher $\fcc = 0.97$ found by \citetalias{weinberg} (vs. 0.82 here) is in better agreement with theoretical expectations that Al is an almost pure CCSN element. Conversely, for K the higher $\fcc=0.94$ (vs. 0.80 in \citetalias{weinberg}) agrees better with the expectation that CCSN dominate its production. For Co we find $\fcc = 0.69$ vs. $\fcc=0.80$ in \citetalias{weinberg}, in better agreement with the \citetalias{griffith19} findings from GALAH but lower than predicted by the yield models of \citet{rybizki17}. For P and V, smaller separation of $\xmg$ values in DR16 leads to higher inferred values of $\fcc$. \citetalias{jonnson20} stress that P and V are two of the least precise and potentially least accurate abundances in DR16. We do not have simple explanations for these differences or a clear indication of which abundances are more accurate for a given element, DR14 or DR16. The enormous sample size and generally high abundance precision in APOGEE allow us to see artifacts that would be hidden in smaller samples or less precise data. The artifacts discussed in Section~\ref{subsec:artifacts} and the two-process parameter differences discussed here imply that systematic uncertainties remain in the APOGEE measurements at a level that is physically interesting.

\subsection{Predicting Bulge Abundances}\label{subsec:predicting}

Using Equations (\ref{eq:aratio}) and (\ref{eq:xmg}) as well as the best-fit two-process model parameters from Table~\ref{tab:zeros}, we can predict the full set of abundance ratios for any star given only its $\mgh$ and $\femg$ values. If the disk and the bulge are enriched by the same nucleosynthetic processes, then the two-process model parameters used to describe the disk should accurately predict the bulge abundances. To test this, we calculate the APOGEE abundance suite of every bulge star using the model parameters from the resampled disk. These predicted distributions in $\xmg$ vs. $\mgh$ space are shown in Figure~\ref{fig:pred_meds}. Median trends of the high-Ia and low-Ia bulge populations (Figure~\ref{fig:XMg}) are overplotted for comparison, including the zero-point offsets discussed above. As we only fit the two-process model above $\mgh = -0.7$, trends below this metallicity are not included in our subsequent comparison and analysis but are included in the plot for completeness.

\begin{figure*}[t]
 \begin{centering}
 \includegraphics[width=.75\textwidth, angle=0]{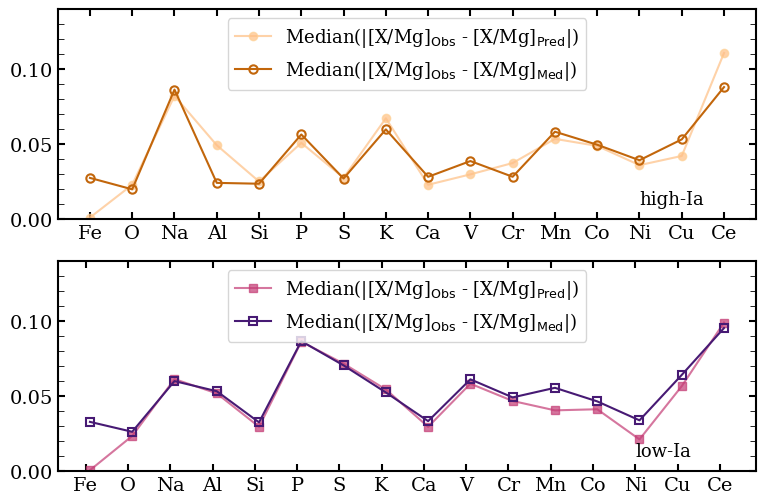}
 \caption{Median absolute difference of the of the observed bulge stars' [X/Mg] (including offsets) and the predicted [X/Mg] values for high-Ia (light orange/top) and low-Ia (pink/bottom) populations. Alongside, we plot the median absolute difference of observed bulge stars' $\xmg$ and the median $\xmg$ value binned by $\mgh$ on the high-Ia (dark orange/top) or low-Ia (purple/bottom) populations. Only stars with Mg/H $>$ -0.7 are included in the calculations.}
 \label{fig:med_comp}
 \end{centering}
\end{figure*}

The $\femg$ vs. $\mgh$ predicted abundances are identical to those observed, by construction. All elements with low $\RIa$ values have `tight' predicted abundance distributions and little spread. This is a direct result of the two-process model assumptions: elements whose production is dominated by CCSN should trace Mg and have little to no separation between the high-Ia and low-Ia sequences. We do \textit{not} add spread induced by observational errors to our predictions, so the width of our predicted abundance distributions is smaller than that of the observed distributions (Figure~\ref{fig:XMg}). The only source of scatter in the two-process predictions comes from the scatter of $\femg$ at fixed $\mgh$, which is interpreted as star-to-star variations in the SNIa-to-CCSN enrichment ratio $\AIa/\Acc$. For elements with $\RIa \approx 0 $ ($\fcc \approx 1$), the induced scatter is minimal because SNIa do not contribute to this element in any case.

Figure~\ref{fig:pred_meds} shows good agreement between the predicted abundances and observed bulge median trends of most elements. The most noticeable exceptions are for Na, Al, P, and Ce, particularly at super-solar metallicities. For Na, this difference reflects the difference in the observed median sequences of the resampled disk and bulge (Figure~\ref{fig:comp_bulge_disk}). For Al, P, and Ce, the power-law form of the two-process model leads to over-predicting the disk trends themselves at high $\mgh$ (Figure~\ref{fig:2proc}), leading to over-prediction of the bulge trends. The trends for many elements diverge from predictions below $\mgh = -0.7$ because the power-law extrapolation of the two-process model becomes inaccurate. 

Interestingly, for [Mn/Mg] the agreement between predicted and observed trends in Figure~\ref{fig:pred_meds} is better than the agreement of the observed medians in Figure~\ref{fig:comp_bulge_disk}. This improvement arises at least partly because the median $\femg$ values are slightly lower for the bulge than for the resampled disk. Within the two-process model, this difference implies that bulge stars have slightly lower SNIa contributions at a given $\mgh$. Because Mn has a large inferred $\RIa = 2.1$, this difference translates into lower predicted [Mn/Mg] for the bulge, producing better agreement. Given the observational uncertainties, these improvements could be a fluke, but it could be a sign that the two-process description is correctly capturing the impact of subtle differences in the relative SNIa/CCSN contributions between the disk and the bulge.

To better quantify the similarities and differences between the observed and predicted populations, we calculate the median of the absolute differences, 
\begin{equation}
    \text{median}(|\xmg_{\rm(obs)} - \xmg_{\rm{pred}}|),
\end{equation}
for each element, including the corrective offsets. In principle the two-process model should be a better predictor of the elemental abundances than the median value of stars on the same sequence with the same $\mgh$ because the model accounts for star-by-star scatter in $\AIa/\Acc$. To test this expectation, we compare the median absolute difference of the observed and the predicted abundances to the median absolute difference of the observed abundance and the median $\xmg$ value for a given $\mgh$ bin,
\begin{equation}
    \text{median}(|\xmg_{\rm{obs}} - \xmg_{\rm{med}}|).
\end{equation}
We plot both statistics for the high-Ia (top) and low-Ia (bottom) populations of all elements in Figure~\ref{fig:med_comp}. We only consider stars with $\mgh > -0.7$, as the two-process model was only fit for such stars. 

To first approximation, Figure 10 shows similar median absolute difference statistics for all elements regardless of whether the two-process model or the observed median sequence is used to predict a star's $\xmg$. This similarity suggests that most of the scatter in $\xmg$ arises from observational errors, and the median absolute differences are indeed largest for elements measured with relatively low precision, such as Na, P, and K. For the high-Ia population, the two-process model predicts [Al/Mg] and [Ca/Mg] worse than the corresponding median trend because it also predicts the median trend itself poorly (see Figure~\ref{fig:pred_meds}). For the low-Ia population, the two-process prediction is more accurate than the observed median trend for Fe-peak elements, particularly for [Mn/Mg] and [Ni/Mg]. This admittedly subtle difference indicates that star-by-star deviations for Fe-peak elements track the SNIa contribution to $\femg$, as the two-process model predicts. Stronger tests of the two-process model can be obtained by focusing on subsets of disk stars with the highest SNR spectra, so that observational contributions to scatter all minimized. We reserve such an investigation for future work.

\section{IMF constraints}\label{sec:IMFs}

Recent works by \citet{ballero2007}, \citet{grieco2012} and \citet{grieco2015} find that the bulge (and galactic center) is better fit by a chemical evolution model with a more top heavy IMF than that which fits the solar neighborhood.  \citet{grieco2012} employ chemical evolution modeling to reproduce the metal rich and metal poor bulge populations. They find that a model with a flatter IMF, such as the \citet{salpeter}, fits the MDF and $\mgfe$ vs. $\feh$ abundance distribution of blue red clump stars from \citet{hill2011}.

Our measurements allow an entirely different test of IMF differences between the bulge and the disk because the relative amounts of different elements produced in a CCSN depend on the mass of the progenitor. Changing the high mass slope of the IMF will increase or decrease the number of massive stars and thus the ratio of nucleosynthetic products from the IMF-integrated CCSN population. If a steeper or shallower IMF induces larger abundance differences than those observed, we can exclude that possible IMF in the bulge.

We adopt a Kroupa IMF \citep{kroupa2001} as our standard for the disk. This three slope IMF is shown in Equation (\ref{eq:IMF}), with $a_3 = -2.3$. To produce a top heavy or light IMF, we change the high mass slope ($a_3$) to -2.0 or -2.6, respectively. Our IMFs are all of the form
\begin{equation}
   \frac{dN}{dM} = 
   \begin{cases}
        A_{\rm K} M^{-0.3}, & M< 0.08 \\
        B_{\rm K} M^{-1.3}, & 0.08 \leq M < 0.5\\
        C_{\rm K} M^{a_3}, & M \geq 0.5
   \end{cases}
   \label{eq:IMF}
\end{equation}
with $A_{\rm K}, B_{\rm K},$ and $C_{\rm K}$ being the appropriate multiplicative constants. 

\begin{figure}[!htb]
 \begin{centering}
 \includegraphics[width=\columnwidth, angle=0]{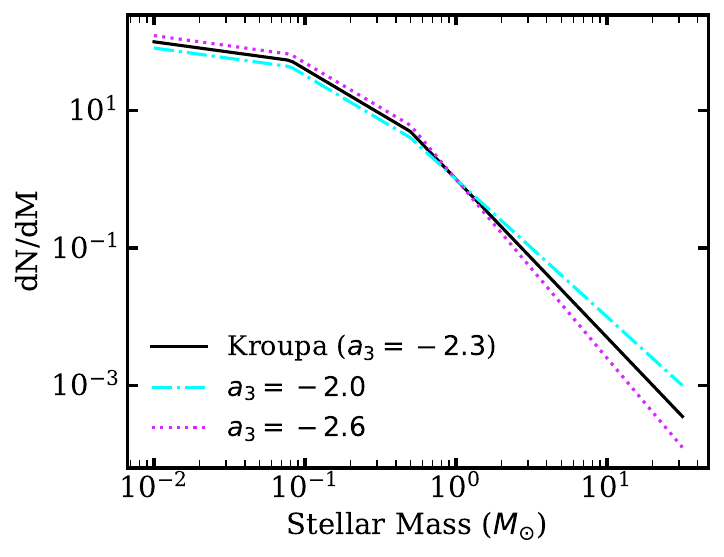}
 \caption{Kroupa IMF (black) and edited Kroupa IMFs with high mass slope of -2.0 (cyan) and -2.6 (pink). We compare the resulting yields of populations integrated under these three IMFs.}
 \label{fig:imfs}
 \end{centering}
\end{figure}

To derive the CCSN abundance yields from our three IMFs, we employ VICE, the Versatile Integrator for Chemical Evolution, and its yield integration function \citep{vice}. In our analysis we take net CCSN yields from \citet[][hereafter CL13]{CL13} at solar metallicity. While \citetalias{CL13} explode all stars to $120\msun$, we assume for our modeling that progenitors above 30$\msun$ collapse to black holes instead of exploding as CCSN. This sharp cutoff is a simplification of an ``explodability landscape'' that is probably much more complex \citep{pejcha2015, tuguldur}. The VICE integrator returns yields in $\msun$ of element X per $\msun$ of stars formed. We convert to bracket notation using solar abundances from \citet{asplund2009}. We note that APOGEE take solar abundance from \citet{grevesse2007}, but we do not expect this to impact our results. 

The resulting yields and [X/Mg] integrated with all three IMFs can be found in Table~\ref{tab:CL13_yield} of Appendix~\ref{ap:IMFs}. This Appendix includes a more detailed exploration of the \citetalias{CL13} yields such as their sensitivity to metallicity and to the choice of upper mass cutoff for explosion. We also include a similar calculation for the \citet[][hereafter LC18]{LC18} yields, which explode stars to 25$\msun$ with different explosion criteria. We note that the \citetalias{CL13} yields force an ejection of $0.1 \msun$ of $^{56}$Ni from all stars.

\begin{figure*}[htb]
 \begin{centering}
 \includegraphics[width=1\textwidth, angle=0]{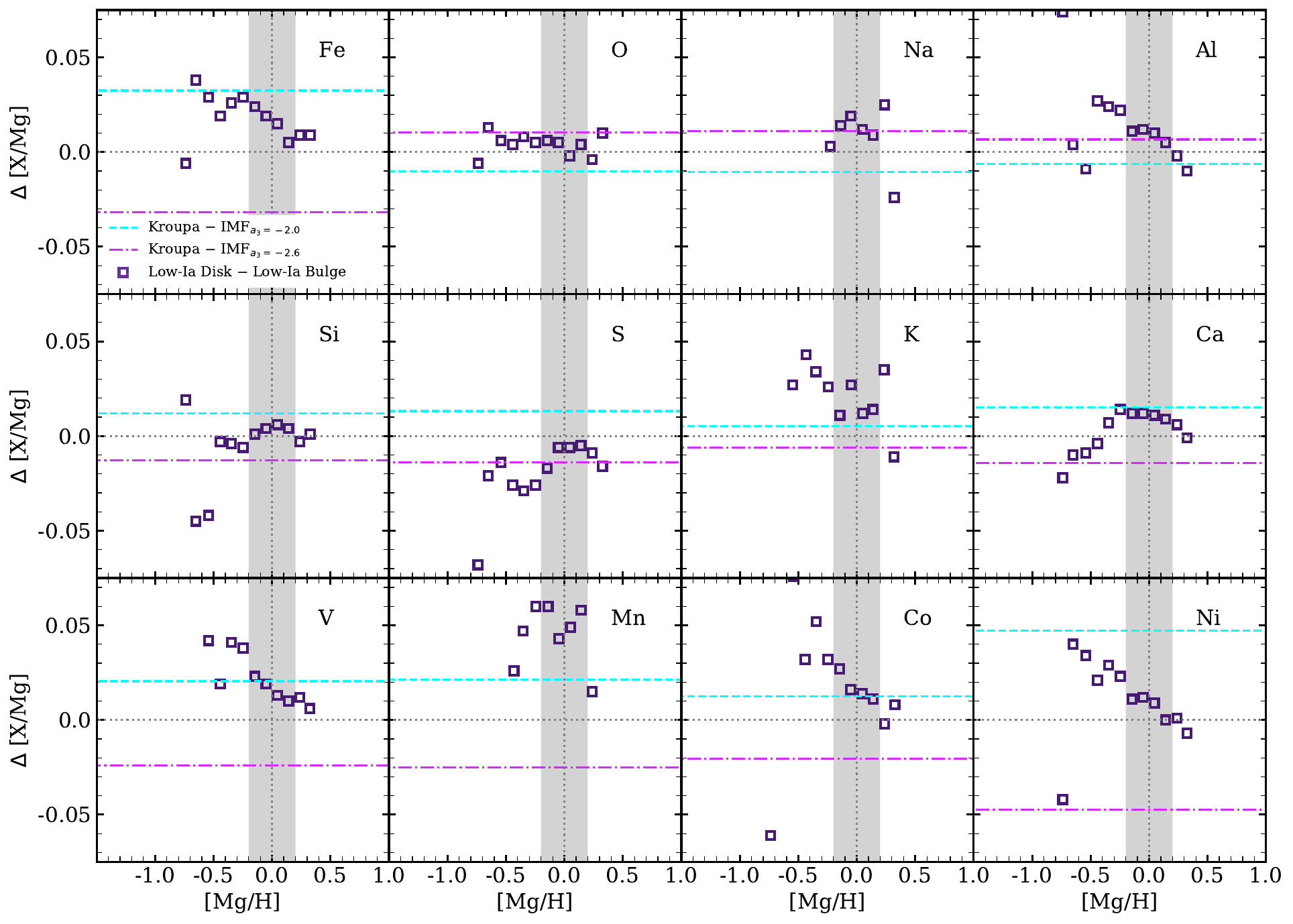}
 \caption{Differences between the bulge and resampled disk low-Ia median sequences (purple squares) and theoretical abundance changes between the standard Kroupa IMF ($a_3 = -2.3$) and a Kroupa IMF with an altered high mass slope. The inducted elemental differences between the Kroupa and $a_3 = -2.0$ IMFs are shown in the cyan dashed lines and those for $a_3 = -2.6$ in the magenta dash-dot lines. The theoretical and observed abundances should be compared near solar $\mgh$, as indicated by the grey band.}
 \label{fig:medDiffsIMF}
 \end{centering}
\end{figure*}

Unfortunately, while predicted yields are mass-dependent, that dependence is fairly similar for all of the elements examined here. As a result, the theoretical [X/Mg] abundances for the Kroupa and $a_3 = -2.0$ or $a_3 = -2.6$ IMFs differ by small amounts, 0.05 dex for [Ni/Mg] and 0.01-0.02 dex for other elements. These IMF-induced abundance differences are plotted in Figure~\ref{fig:medDiffsIMF}. Taking Kroupa as the standard, we show the $\Delta \xmg$ values for 12 elements, where
\begin{equation}
    \Delta \xmg = \xmg_{\rm Kroupa} - \xmg_{a_3 = -2.0}
\end{equation}
or 
\begin{equation}
    \Delta \xmg = \xmg_{\rm Kroupa} - \xmg_{a_3 = -2.6}.
\end{equation}
Although these values are plotted as horizontal lines, they are calculated at solar $\feh$. We stress that this calculation only investigates star formation changes that would affect the CCSN contribution to each element. We do not account for any differences in SNIa nucleosynthesis between the bulge and disk at fixed $\mgh$ and $\femg$. A difference in SNIa-produced $\xfe$ ratios could compensate for, or masquerade as, an IMF-induced change to CCSN contributions. 

For comparison, Figure\ref{fig:medDiffsIMF} also shows the observed differences in the bulge and resampled disk low-Ia median points, 
\begin{equation}
    \Delta \xmg = \xmg_{\rm Disk} - \xmg_{\rm Bulge}.
\end{equation}
We compare only the low-Ia median abundances as the bulge low-Ia sequence is more well populated than the high-Ia sequence, and we wish to minimize the SNIa contribution for this test in any case. We have already seen in Figure~\ref{fig:comp_bulge_disk} that disk and bulge median sequences agree well, and in Figure~\ref{fig:medDiffsIMF} we see that the differences near solar metallicity are typically 0.02 dex or below. These are comparable in magnitude to the predicted IMF effects, but some elements agree better with the steeper IMF and some with the shallower IMF, with no clear pattern to separate these groups. We are therefore inclined to ascribe these differences to residual systematics in matching disk and bulge abundance scales at the 0.02-dex level. 

Two elements deserve special mention. [Mn/Mg] ratios show the largest differences between disk and bulge, $\sim$0.05 dex, much larger than predicted by an IMF change. If real, this difference likely arises from different SNIa enrichment in the bulge, perhaps connected to the strong apparent metallicity dependence of Mn yields. [Ni/Mg] ratios, on the other hand, show 0.01-dex agreement near solar metallicity, much closer than the 0.04-dex changes predicted by the IMF models. At face value this comparison implies a stringent limit on IMF differences, with $|\Delta a_3| \lesssim 0.1$. However, Ni predictions may be sensitive to the criteria used to define the mass cut in supernova models. Ideally we would like to base an IMF test on relative abundances of Mg, O, Si, and Ca, which are well measured and are predicted (and empirically inferred) to come predominantly from CCSN with little metallicity dependence. If abundance scales of different populations can be reliably cross-calibrated at the 0.01 dex level, then tests of IMF variations at the $|\Delta a_3| \approx 0.2-0.3$ level can be achieved.

\section{Summary}\label{sec:summary}

Using APOGEE DR16+ data, we present the $\xmg$ vs. $\mgh$ and $\xfe$ vs. $\feh$ abundance trends of the Milky Way bulge ($R<3 \kpc$) for Fe, O, Na, Al, Si, P, S, K, Ca, V, Cr, Mn, Co, Ni, Cu, and Ce, the latter two being new to DR16. The addition of the LCO instrument in APOGEE DR16 provides us with observations of stars in the southern hemisphere and thus a more complete look at the bulge than previous APOGEE data releases. After a series of quality cuts, we are left with 11,229 bulge stars with median SNR of 149. 

Like the disk, the bulge population shows a bimodal $\mgfe$ vs. $\feh$ distribution. The DR16+ data reveal that the two populations are distinct and do not form a continuous sequence. As in \citetalias{weinberg}, we divide the sample into high-Ia and low-Ia populations in this plane. We see median $\mgfe$ vs. $\feh$ trends that match those of the \citetalias{weinberg} disk definition, though slightly offset. As in \citet{hayden2015} and \citetalias{weinberg}, we find that the inner Galaxy contains more low-Ia stars than high-Ia stars, with the stellar distribution extending to $\feh$ of -1.5.

In Section~\ref{sec:abundnaces} we present the $\xmg$ vs. $\mgh$ and the $\xfe$ vs. $\feh$ abundance distributions and $\xmg$ vs. $\mgh$ median trends for the APOGEE elements. Some data systematics can be seen, including a finger-like feature in low-Ia O, Ca, and Si in $\xfe$ space; a bimodal distribution of high-Ia Cr stars; and banding/clumping in the Al, P, and V abundance distributions. We only exclude the lower [Cr/Mg] high-Ia stars in our analysis, but we explore the impact of these systematics in Appendix~\ref{ap:systematics}. 

The main goal of this paper is to determine the similarity or dissimilarity between the median high-Ia and low-Ia $\xmg$ trends of the Galactic bulge and disk. However, upon a comparison of our bulge sample with the disk as defined by \citetalias{weinberg}, we find systematic abundance differences which correlate with the stellar $\logg$ distribution. We therefore resample the APOGEE DR16+ disk to select a subset of stars that reproduce the $\logg$ distribution of the bulge. We divide the resampled disk stars into high-Ia and low-Ia populations. 

The similarities between the median high-Ia and low-Ia trends of the bulge and disk shown in Figure~\ref{fig:comp_bulge_disk} are striking. All elemental trends agree to within 0.1 dex and most within 0.05 dex; near solar $\mgh$ the agreement of the low-Ia median trends is typically 0.01-0.02 dex (Figure~\ref{fig:medDiffsIMF}). The [Na/Mg] and [Mn/Mg] vs. [Mg/H] bulge high-Ia median trends rise more steeply than those of the disk. The low-Ia [Mn/Mg] median trend is about 0.05-dex higher in the bulge than the disk. The Co bulge medians are offset to lower [Co/Mg] than the resampled disk at low $\mgh$ These differences could reflect differences in the chemical evolution of the disk and bulge, but they are small enough that they could still reflect imperfect relative calibration of bulge and disk abundance scales.

This close agreement of median $\xmg$ trends between the bulge and disk extends the universality of these trends as a function of $R$ and $|Z|$ within the disk found by \citetalias{weinberg}. It demonstrates that although the distribution of stars in $\mgh$ and $\femg$ depends strongly on location in the Galaxy, the physics that determines $\xmg$ ratios for a given $\mgh$ and $\femg$ does not. To quantify this point we compare the observed bulge abundance ratios with those predicted by \citetalias{weinberg}'s empirical two-process model fit to the resampled disk. The two-process model describes the abundances of a star or population of stars as the sum of prompt (CCSN) and delayed (SNIa) components, with each of these contributions to (X/Mg) having a power law dependence. Using the parameters derived as the best fit to the resampled disk's median abundances for each element ($\RIa$, $\acc$, and $\aIa$), we predict the bulge elemental abundances suite based only on the observed [Fe/Mg] and $\mgh$ of the bulge stars. Figure~\ref{fig:pred_meds} shows that the predicted abundances accurately trace the observed [X/Mg] median trends. Elements for which the predicted and observed trends diverge show poor two-process model fits (e.g. Al) and/or real differences in the median trends of the bulge and resampled disk (e.g. Na, Mn, Co). Star-by-star deviations from median trends appear to be dominated by observational errors, but the two-process model explains a small fraction of the scatter for some iron peak elements.

As prior works have supported a bulge IMF with a shallower high mass slope than the solar neighbourhood \citep{grieco2012}, we test if the bulge and sampled disk's similar abundance tracks can constrain the Galactic IMF variability. Using the yield integration function of VICE \citep{vice}, we calculate the theoretical [X/Mg] abundance at solar $\feh$ for the \citetalias{CL13} yields with three IMFs: Kroupa with high mass slopes of -2.0, -2.3, or -2.6. Overall, we find small predicted [X/Mg] changes with a varying IMF, $<0.05$ dex for all elements. If the bulge and disk have different IMFs, we would expect the median abundance differences of the bulge and resampled disk to correlate with the theoretical abundance changes induced by an altered IMF. However, we do not see consistent agreement between the observed and theoretical $\Delta \xmg$ trends for either IMF variation (Figure~\ref{fig:medDiffsIMF}). The observed agreement of median trends argues against IMF slope differences larger than about 0.3. At face value the agreement for [Ni/Mg] implies a more stringent limit, but the predicted sensitivity of Ni yields to the IMF may not be robust. The 0.05 dex difference in [Mn/Mg] trends could be a sign of differences in SNIa enrichment between bulge and disk.

Our principal finding is that the bulge and resampled disk have similar [X/Mg] vs. $\mgh$ median abundance tracks, extending the conclusions of \citet{hayden2015} and \citetalias{weinberg} to the inner Galaxy. The universality of the median [X/Mg] abundance trends suggests that they are not sensitive to most aspects of chemical evolution and instead depend on the IMF-averaged nucleosynthetic yields, which appear consistent throughout the Galaxy. As we obtain additional high SNR observations of the Milky Way bulge, our understanding of the Galactic nucleosynthetic processes will grow. In future works  with APOGEE and other data sets, we aim to set more stringent empirical constraints on the astrophysics that governs the creation of the elements throughout the Galaxy and its closest neighbors.

\section*{Acknowledgements}

We thank APOGEE collaborators Julie Imig, Jianhui Lian, Steve Majewski, Matthew Shetrone, and Gail Zasowski, who were in attendance at the Autumn 2019 OSU workshop and Autumn 2019 bulge workshop and provided many insights on APOGEE abundances and findings for the bulge. We thank the Center for Cosmology and AstroParticle Physics for helping to organize the Autumn 2019 OSU workshop. 

This work was supported by NSF grant AST-1909841. 
D.A.G.H. acknowledges support from the State Research Agency (AEI) of the Spanish Ministry of Science, Innovation and Universities (MCIU), and the European Regional Development Fund (FEDER) under grant AYA2017-88254-P.
S.H. is supported by an NSF Astronomy and Astrophysics Postdoctoral Fellowship under award AST-1801940. 
H.J. acknowledges support from the Crafoord Foundation, Stiftelsen Olle Engkvist Byggm\"astare, and Ruth och Nils-Erik Stenb\"acks stiftelse.
D.M.N. acknowledges support from NASA under award Number 80NSSC19K0589. 
A.R-L acknowledges financial support provided in Chile by Comisi\'on Nacional de Investigaci\'on Cient\'ifica y Tecnol\'ogica (CONICYT) through the FONDECYT project 1170476 and by the QUIMAL project 130001.

Funding for the Sloan Digital Sky Survey IV has been provided by the Alfred P. Sloan Foundation, the U.S. Department of Energy Office of Science, and the Participating Institutions. SDSS-IV acknowledges
support and resources from the Center for High-Performance Computing at
the University of Utah. The SDSS web site is www.sdss.org.

SDSS-IV is managed by the Astrophysical Research Consortium for the 
Participating Institutions of the SDSS Collaboration including the 
Brazilian Participation Group, the Carnegie Institution for Science, 
Carnegie Mellon University, the Chilean Participation Group, the French Participation Group, Harvard-Smithsonian Center for Astrophysics, 
Instituto de Astrof\'isica de Canarias, The Johns Hopkins University, Kavli Institute for the Physics and Mathematics of the Universe (IPMU) / 
University of Tokyo, the Korean Participation Group, Lawrence Berkeley National Laboratory, 
Leibniz Institut f\"ur Astrophysik Potsdam (AIP),  
Max-Planck-Institut f\"ur Astronomie (MPIA Heidelberg), 
Max-Planck-Institut f\"ur Astrophysik (MPA Garching), 
Max-Planck-Institut f\"ur Extraterrestrische Physik (MPE), 
National Astronomical Observatories of China, New Mexico State University, 
New York University, University of Notre Dame, 
Observat\'ario Nacional / MCTI, The Ohio State University, 
Pennsylvania State University, Shanghai Astronomical Observatory, 
United Kingdom Participation Group,
Universidad Nacional Aut\'onoma de M\'exico, University of Arizona, 
University of Colorado Boulder, University of Oxford, University of Portsmouth, 
University of Utah, University of Virginia, University of Washington, University of Wisconsin, 
Vanderbilt University, and Yale University.

This research made use of Astropy,\footnote{http://www.astropy.org} a community-developed core Python package for Astronomy \citep{astropy2013, astropy2018}.

\software{VICE \citep{vice}, Astropy \citep{astropy2013, astropy2018}}

\appendix
\section{The effect of data systematics on median trends}\label{ap:systematics}

In Section~\ref{sec:abundnaces}, we present the median high-Ia and low-Ia abundance trends for APOGEE elements and discuss some of the systematics and artifacts afflicting the data in Section~\ref{subsec:artifacts}. In the $\xfe$ vs. $\feh$ plots (Figure~\ref{fig:XFe}) we see a finger-like feature in O, Ca, and Si; bimodality in the Cr high-Ia stars; banding in Al and P; and a clump in V. In our data analysis we exclude some stars to mitigate the Cr feature and proceed with the other artifacts included. Here, we take a closer look these anomalies to see how the median trends change with their inclusion and exclusion. 

\begin{figure*}[!htb]
 \begin{centering}
 \includegraphics[width=1\textwidth, angle=0]{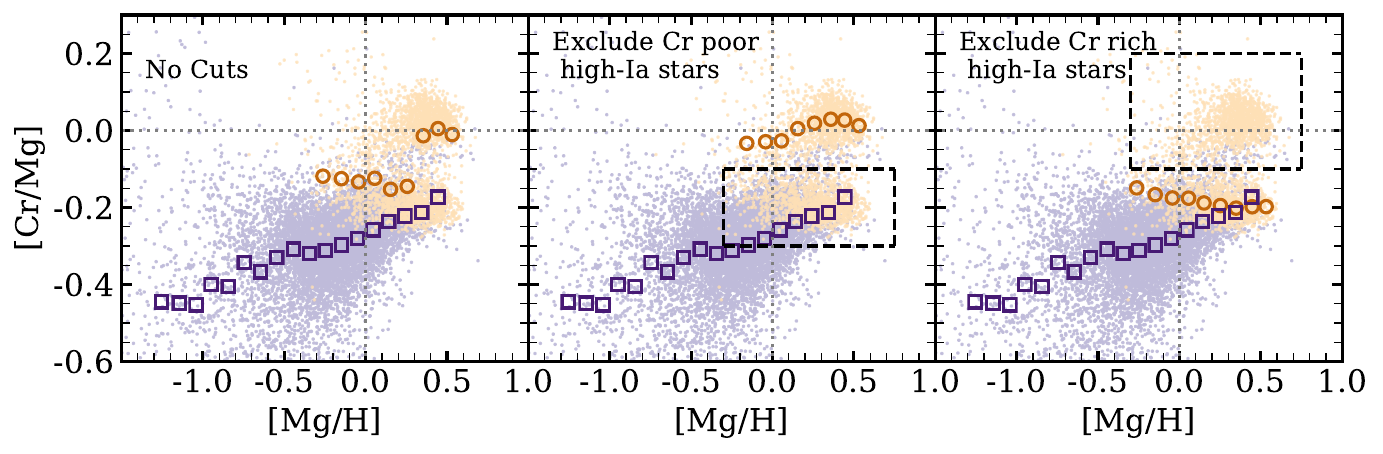}
 \caption{Bulge [Cr/Mg] vs. [Mg/H] distribution with high-Ia stars in orange and low-Ia stars in purple. The medians of the high/low-Ia populations are over-plotted in dark orange/purple circles/squares, where we bin by 0.1 dex and require $>20$ stars per bin. Left: Median trends for the full population. Middle: Median trends excluding the Cr poor high-Ia stars inside the dashed box (exclusion employed in the main body of the paper). Right: Median trends excluding the Cr rich high-Ia stars inside the dashed box.}
 \label{fig:ApA_Cr}
 \end{centering}
\end{figure*}

Figure~\ref{fig:ApA_Cr} presents three versions of the [Cr/Mg] median high-Ia trend. The first panel shows the observed distribution with no cuts. The high-Ia median trend follows the Cr-poor stars at low $\mgh$ but jumps to solar [Cr/Mg] around $\mgh \sim 0.3$. The bimodality in the Cr abundance obviously produces a skewed median trend. In our analysis, we remove 1421 stars with $0 < \mgh < 0.75$ and $-0.3<[\rm Cr/ \rm Mg] < -0.1$. This exclusion and the resulting median high-Ia trend are shown in the middle panel of Figure~\ref{fig:ApA_Cr}. Alternatively, we could have excluded the 1544 Cr-rich stars with $0 < \mgh < 0.75$ and $-0.1<[\rm Cr/ \rm Mg] < 0.2$, as shown in the right panel. As expected, the high-Ia median trend changes dramatically between the two exclusions. The higher $\logg$ disk better resembles the the Cr-rich high-Ia stars and previous works, so we use this subset in our analysis.

\begin{figure*}[!htb]
 \begin{centering}
 \includegraphics[width=0.9\textwidth, angle=0]{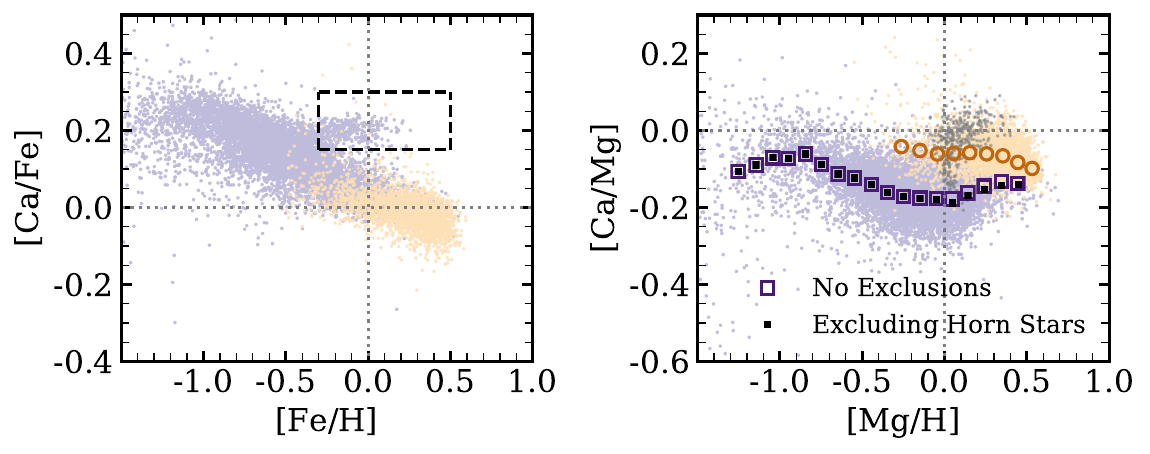}
 \caption{Left: Bulge [Ca/Fe] vs. [Fe/H] distribution with high-Ia stars in orange and low-Ia stars in purple. The over-plotted dashed box identifies the finger stars. Right: [Ca/Mg] vs. [Mg/H] distribution. The medians of the full high/low-Ia populations are over-plotted in dark orange/purple circles/squares, where we bin by 0.1 dex and require $>$20 stars per bin. The black squares show the resulting low-Ia median trend when we exclude the finger stars (shown in grey).}
 \label{fig:ApA_Ca}
 \end{centering}
\end{figure*}

\begin{figure*}[!htb]
 \begin{centering}
 \includegraphics[width=0.9\textwidth, angle=0]{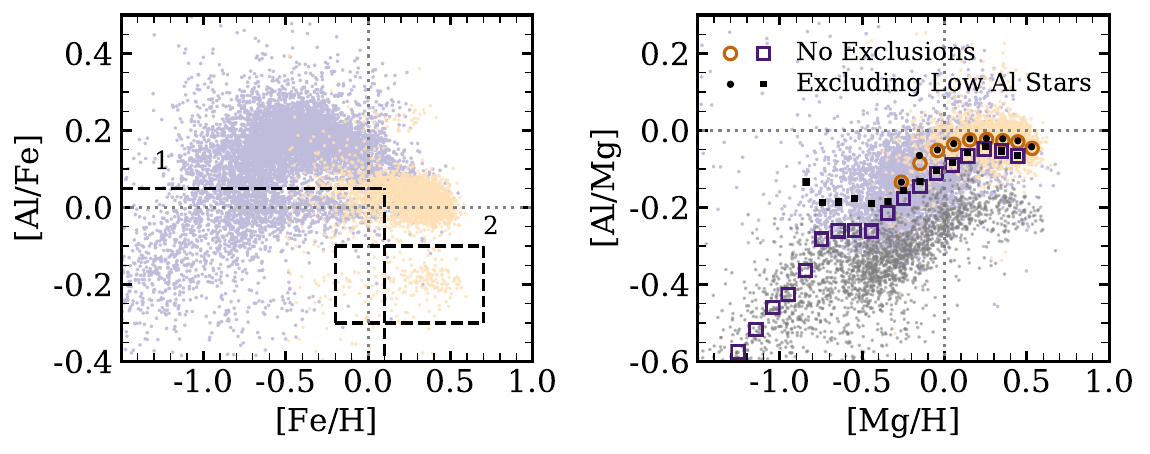}
 \caption{Same as Figure~\ref{fig:ApA_Ca}, but for Al. In the left panel, exclusion region 1 (2) applies to the low-Ia (high-Ia) stars. All excluded stars are shown in grey in the right panel.}
 \label{fig:ApA_Al}
 \end{centering}
\end{figure*}

While we mitigate the Cr systematic with exclusions in our data analysis, the $\alpha$ finger and other banding/clumping are left in. Figure~\ref{fig:ApA_Ca} and~\ref{fig:ApA_Al} show examples of how the [Ca/Mg] and [Al/Mg] vs. [Mg/H] trends (respectively) would change if we removed these features. For Ca, we isolate the finger stars in [Ca/Fe] vs. [Fe/H] space, where the finger is much easier to identify. We define the finger as low-Ia stars with $-0.3 < \feh < 0.5$ and $0.15<[\rm Cr/ \rm Fe] < 0.3$. Median trends including and excluding these stars are plotted in the right panel of Figure~\ref{fig:ApA_Ca}. The low-Ia median trend changes insignificantly after the exclusion of 368 finger stars. We find that the inclusion of the bulge finger stars in our O, Si, and Ca analysis does not affect our resulting median trends or conclusions. 

The low abundance trends \citepalias{jonnson20} seen in Al and P are much harder to isolate as they blend into the main abundance tracks. The left panel of Figure~\ref{fig:ApA_Al} plots [Al/Fe] vs. [Fe/H], where the banding is more pronounced than in Mg space. We define two exclusion regions. The first (1) is [Al/Fe]$<0.05$ and [Fe/H]$<0.1$ for low-Ia stars and the second (2) is $-0.1<[\rm Al/\rm Fe] < -0.3$ and $-0.2< \feh < 0.7$ for high-Ia stars. Both boxes are drawn on the figure and include 2700 and 206 stars, respectively. Median trends including and excluding these stars are shown in the right panel.  The high-Ia trend is not noticeably affected by the small exclusion, but the low-Ia trend deviates from the full population median at low $\mgh$. This deviation should be expected, as we cut out many of the low $\mgh$ stars. While the Al `tail' may be a systematic, the median trends above $\mgh = -0.25$ are robust. As the bulge and disk median trends agree at this metallicity, our conclusions about the similarity of the bulge and disk abundance ratios hold.

\begin{figure*}[!htb]
 \begin{centering}
 \includegraphics[width=\textwidth, angle=0]{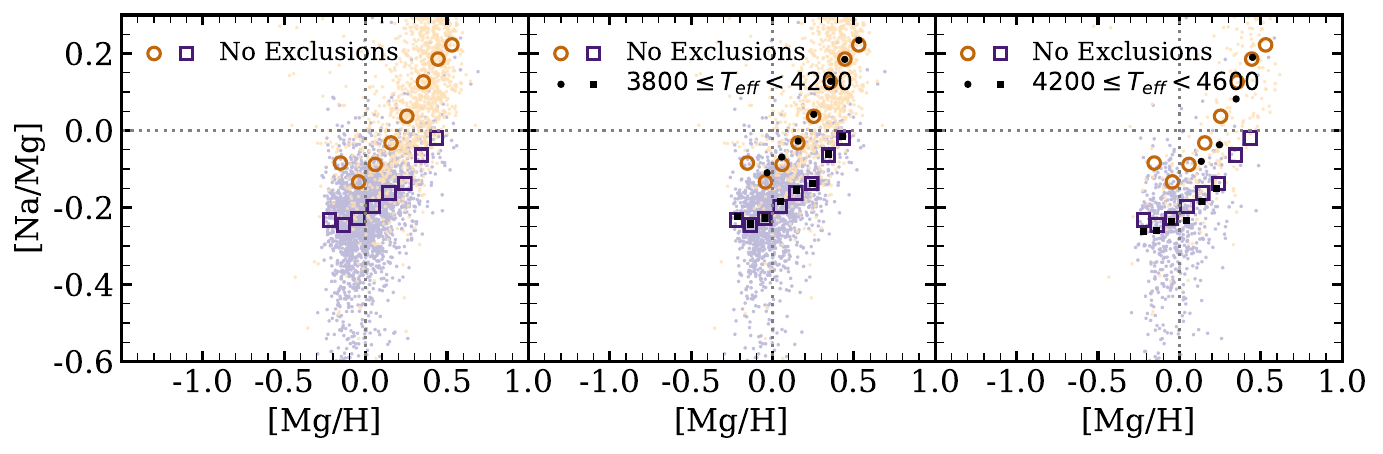}
 \caption{Bulge [Na/Mg] vs. [Mg/H] distribution with high-Ia stars in light orange and low-Ia stars in light purple. The medians of the full high/low-Ia populations are over-plotted in dark orange/purple circles/squares, where we bin by 0.1 dex and require $>20$ stars per bin. Left: Full sample (3851 stars) These medians remain the same in each panel. Center: Medians of stars with $3800 \kel \leq \teff < 4200 \kel$ (2800 stars). Medians of the full population are in orange/purple and of the temperature cut sample are in black. The background light orange/purple points show the sub-sample of stars in  the temperature range.  Right: Same as middle panel, but for stars with $4200 \kel \leq \teff < 4600 \kel$ (931 stars) in black.}
 \label{fig:ApA_Na}
 \end{centering}
\end{figure*}

We also explore the temperature correlation of median abundance trends. As noted in Section~\ref{sec:abundnaces}, APOGEE flags low temperature Na, K, and Mn stars due to reduction systematics that correlate with temperature. To see if this affects the median high-Ia and low-Ia trends we compare the full sample medians to those of smaller temperature divisions. Figure~\ref{fig:ApA_Na} plots [Na/Mg] vs. [Mg/H] distributions and medians for a sample with no temperature cuts, stars with $3800 \kel \leq \teff < 4200 \kel$ (2800), and stars with $4200 \kel \leq \teff < 4600 \kel$ (931). The majority of stars lie within the lower temperature range, so the medians of this sample track the full population. Both  median sequences of the higher temperature sample lie below those of the full population, though the differences are $<-0.05$ dex. The median high-Ia and low-Ia Na trends appear to have some small correlation with temperature. 

\begin{figure*}[!htb]
 \begin{centering}
 \includegraphics[width=\textwidth, angle=0]{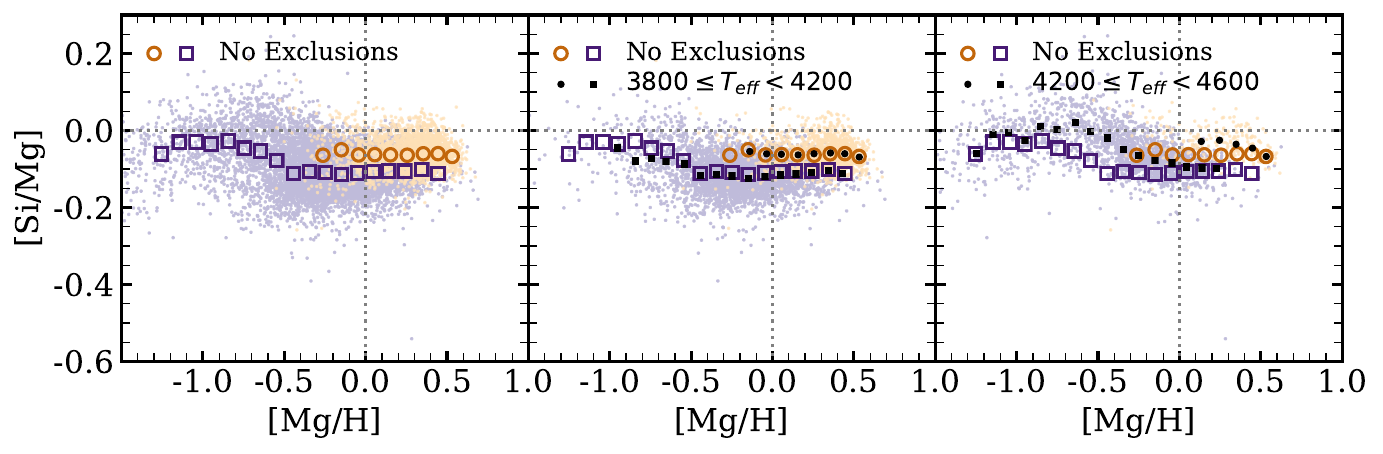}
 \caption{Same as Figure~\ref{fig:ApA_Na}, but for Si. There are 5254 stars in the lower temperature range and 2208 in the higher temperature range.}
 \label{fig:ApA_Si}
 \end{centering}
\end{figure*}

We repeat this analysis with all other elements using the same temperature divisions. In most cases, the lower temperature bin dominates the sample and agrees well with the full population medians. We see no temperature dependent changes in the $\xmg$ vs $\mgh$ median trend location for Ca, K, Mn, Ni, Cu, and Ce. In the higher temperature bin, we find that the low-Ia trends of Al, O, and Si sit slightly above the full population medians and those of Co and Cr sit below (in many cases there are too few high-Ia median points to draw strong conclusions). Si displays one of the largest low-Ia median trend changes in the higher temperature bin ($\sim0.1$ dex), shown in Figure~\ref{fig:ApA_Si}. Some elemental abundance variations with temperature, such as Si, are on the scale of or larger than the difference between the bulge and disk medians. 

While we match $\logg$ distributions between the bulge and disk samples, we do not match $\teff$ distributions. To verify that our conclusions hold, we check the median high-Ia and low-Ia disk trends' variation with temperature. We find similar deviations in the disk as in the bulge, e.g. the median Si high-Ia and low-Ia trends of stars with $4200 \kel \leq \teff < 4600 \kel$ in the disk lie above those of the full population by a similar amount to that seen in Figure~\ref{fig:ApA_Si}. While there is uncertainty in the exact location of the abundance trends, we have confidence that the bulge and the disk trends agree. 

\section{Cu and Ce abundances} \label{ap:Ce_Cu}

\begin{figure}[!htb]
 \begin{centering}
 \includegraphics[width=0.9\textwidth, angle=0]{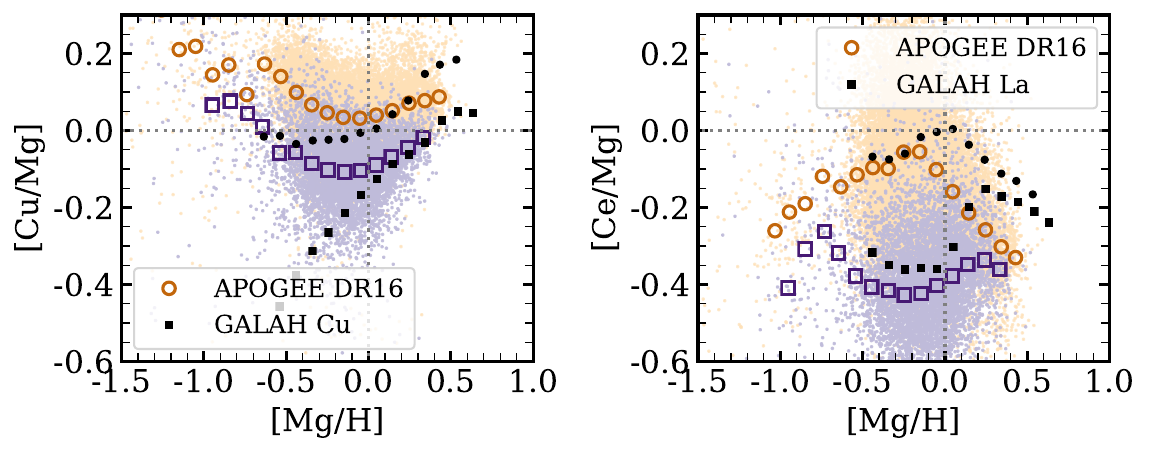}
 \caption{APOGEE [Cu/Mg] (left) and [Ce/Mg] (right) vs. [Mg/H] abundances for the W19 disk, with high-Ia stars in light orange and low-Ia stars in purple. The medians of the high-Ia and low-Ia populations are over plotted in dark orange circles and dark purple squares, respectively. GALAH high-Ia (black circles) and low-Ia (black squares) medians from \citetalias{griffith19} are also included for [Cu/Mg] (left) and [La/Mg] (right) \citepalias{griffith19}.}
 \label{fig:Ap_B_abund}
 \end{centering}
\end{figure}

While the main body of this paper focuses on the abundance trends in the Milky Way bulge, in this appendix we will take a closer look at the disk Cu and Ce abundances, continuing the work of \citetalias{weinberg} and \citetalias{griffith19} for the two new APOGEE elements. In Figure~\ref{fig:Ap_B_abund} we plot the Cu and Ce abundances for all stars using the \citetalias{weinberg} disk sample cuts ($1 < \logg < 2$, $3700 \kel < \teff < 4600 \kel$,  $3 \kpc < R < 15 \kpc$, $|Z| < 2 \kpc$) with median high-Ia and low-Ia trends (calculated as in Section~\ref{sec:abundnaces}). Over plotted are the GALAH median high-Ia and low-Ia trends for Cu and La from \citetalias{griffith19}.

APOGEE and GALAH Cu abundances agree reasonably well above solar $\mgh$. Here, both surveys' [Cu/Mg] median trends show large sequence separation and positive slopes. The GALAH medians are significantly more inclined, especially when comparing the high-Ia sequences. As noted in Section~\ref{subsec:comparison}, the APOGEE Cu abundances turn upwards towards higher [Cu/Mg] at low [Mg/H]. We do not trust the measurements for low $\mgh$ stars.

As GALAH does not report Ce abundances, we first compare Ce to La, the closest observed element on the periodic table with similar neutron capture origins. Both elements show the expected non-linear trends. In Ce and La we see a clear peak in the high-Ia median trends near solar. The same is seen in the low-Ia medians, though the Ce peak is smaller and offset to higher $\mgh$ (we ignore the upturned low-Ia tail at low $\mgh$ as there are fewer stars here). As explained in \citetalias{griffith19}, the trends rise with $\mgh$ in the low metallicity regime because of the increasing numbers of Fe seeds and then decline with $\mgh$ in the high metallicity regime because of the decreasing neutron-to-seed ratio \citep{gallino98}.

\add{We further compare APOGEE Ce abundances with other high resolution, high SNR studies from \citet{battistini2016} (361 dwarf stars in the solar neighborhood) and \citet{delgado2017} (1012 dwarf stars from the HARPS-GTO survey). We divide the high-Ia and low-Ia stars according to Equation~\ref{eq:boundary} and bin median points by 0.1 dex in [Mg/H]. We see agreement between APOGEE and both studies. The smaller data set from \citet{battistini2016} shows a declining high-Ia trend and a flat low-Ia trend near solar, in agreement with APOGEE data. The \citet{delgado2017} Ce abundances also appear in agreement with APOGEE, though the trends are offset to higher [Ce/Mg]. Neither study include sufficient high-Ia stars with [Mg/H] $< 0.04$ to identify the high-Ia peak seen in APOGEE. We also refer the reader to \citet[][Figure 14]{jonnson20}, who compare APOGEE DR16 Ce abundances with those of high-resolution bulge giants from \citet{forsberg2019} and find reasonable agreement between the two surveys. Conclusions from APOGEE's Ce trends, however, should be drawn with caution, as they are derived from a single line and prone to large uncertainties. 
}

\begin{figure}[!htb]
 \begin{centering}
 \includegraphics[width=1\textwidth, angle=0]{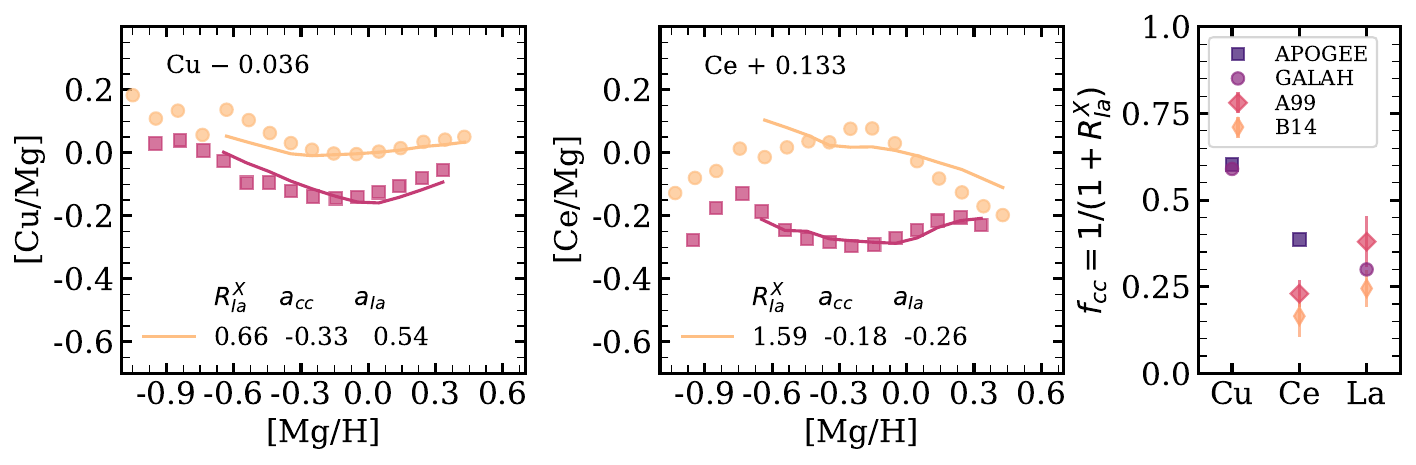}
 \caption{Left: Same as Figure~\ref{fig:2proc} but for Cu abundances in a sample with the \citetalias{weinberg} disk cuts. Center: Same for Ce abundances. Right: Fractional CCSN contribution to Cu, Ce, and La as calculated from $\RIa$ values for fits to APOGEE (dark purple squares) and GALAH (light purple circles). Theoretical $r$-process contributions from \citet{arlandini99}(A99, pink diamonds) and \citet{bisterzo14}(B14, thin orange diamonds) are included for Ce and La.}
 \label{fig:Ap_B_2proc}
 \end{centering}
\end{figure}

We fit both the [Cu/Mg] and [Ce/Mg] disk median trends with the two-process model, as outlined in Section~\ref{sec:two-proc} (Figure~\ref{fig:Ap_B_2proc}).  As both elemental trends are non-linear, the two-process model does not reproduce observations at the low metallicity end. However, it does give us a sense of the sequence separation at $\mgh = 0$. \citetalias{griffith19} find Cu to have an $\RIa = 0.71$, $\acc = 0.56$, and $\aIa=-0.40$. The $\acc$ and $\aIa$ values found for APOGEE differ drastically (likely due to the upturn at low metallicity) but the $\RIa$ value of 0.66 is in good agreement with GALAH. APOGEE's $\RIa$ for Ce and GALAH's La differ more, at 1.59 and 2.31, respectively. The $\acc$ and $\aIa$ terms hold less meaning, as the abundance tracks do not follow a power law dependence. 

In the right hand panel of Figure~\ref{fig:Ap_B_2proc}, we plot the $\fcc$ values corresponding to all four $\RIa$ measurements. The GALAH and APOGEE points for Cu overlap around $\fcc = 60 \%$, showing strong agreement that Cu has a large delayed component. As detailed in \citetalias{griffith19}, this component may be due to AGB stars, as SNIa models do not produce substantial amounts of Cu. For Ce and La, the $\fcc$ value is better interpreted as a fractional $r$-process component ($f_r$). APOGEE and GALAH find Ce and La, respectively, to have around 30-40\% $r$-process contribution. We note that our two-process models are scaled to CCSN and SNIa, so this estimated percentage could change if the $s$-process and SNIa enrichment have different delay times. The $f_r$ values for our neutron capture elements agree reasonably well with theory, which predicts both to be $s$-process dominated. \citet{arlandini99} find Ce to have 77\% $s$-process contribution ($f_s$) and \citet{bisterzo14} find 83\%. See \citetalias{griffith19} for a broader discussion of La. Figure~\ref{fig:Ap_B_2proc} plots these theoretical $f_r$ value for both elements, taking $f_r = 1-f_s$. 

\section{CCSN yield exploration} \label{ap:IMFs}

In Section~\ref{sec:IMFs}, we present the theoretical [X/Mg] abundances at solar [Fe/H], as calculated from the net \citetalias{CL13} yields integrated with three different IMF high mass slopes. In this appendix, we provide additional information about the net \citetalias{CL13} and \citetalias{LC18} CCSN yields and their resulting [X/Mg] abundances. While \citetalias{LC18} cite multiple improvements to the \citetalias{CL13} approach, we note differences in their explosion criterion: \citetalias{CL13} impose a mass cut that ejects 0.1 $\msun$ of $^{56}$Ni for all stars while \citetalias{LC18} set interior [Ni/Fe] values and choose a mass cut that ejects 0.07 $\msun$ of $^{56}$Ni for stars with $M\leq 25\msun$. Both compute yield calculations for rotating stars, but we do not explore rotation effects here.

\begin{figure}[!htb]
 \begin{centering}
 \includegraphics[width=0.9\columnwidth, angle=0]{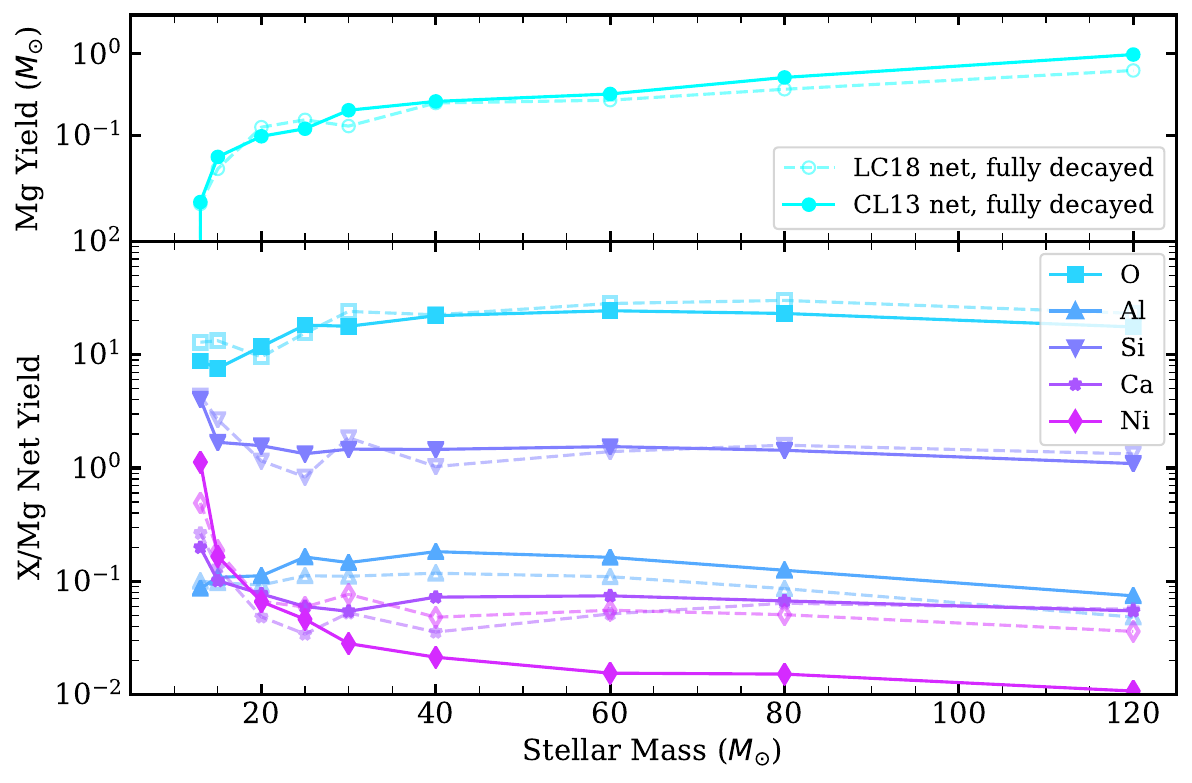}
 \caption{Elemental net yields from \citetalias{CL13} (solid) and \citetalias{LC18} (dashed) as a function of mass for $\feh=0$ with no rotation. Top: Net explosive and wind Mg yields in $\msun$ (cyan circles) for published CL13 (solid marker and line) and unpublished LC18 (empty circle and dashed line). Both are reported after a full decay of unstable isotopes. Bottom: Net X/Mg yields for O (light blue squares), Al (dark blue upwards triangle), Si (indigo downwards triangles), Ca (purple stars), and Ni (pink diamonds) for both CL13 and LC18.}
 
 \label{fig:ApC_mass_yield}
 \end{centering}
\end{figure}

The foundation of our earlier IMF investigation rests on the fact that CCSN yields have a mass dependence. If the high mass IMF slope steepens/flattens, then the Galaxy sees fewer/more high mass stars and their nucleosynthetic products. The IMF integrated abundances shown above are thus dependent upon the elemental yield's mass dependence. In our case, we need to know how the yields vary relative to Mg. Figure ~\ref{fig:ApC_mass_yield} shows the net Mg yield and the net X/Mg yield for O, Al, Si, Ca, and Ni as a function of mass for \citetalias{CL13} and \citetalias{LC18}. In the \citetalias{CL13} yields we see that while Mg has an obvious mass dependence, the alpha elements (O, Si, Ca) and light odd-$Z$ element (Al) track Mg and thus have little mass dependence in X/Mg. This explains our findings in Figure~\ref{fig:medDiffsIMF}, where all four elements show small $\Delta \xmg$ for a changing IMF. The Ni/Mg yields steeply decrease with stellar mass, causing larger changes in the [Ni/Mg] values when the IMF is varied. 

While \citetalias{CL13} report yields for all massive stars, \citetalias{LC18} publish their set R yields, where stars above $25\msun$ are assumed to collapse to a black hole. Figure~\ref{fig:ApC_mass_yield} plots their M set, which imposes the same mass cut as set R, but explodes stars to 120 $\msun$. These yields show very similar mass dependence to \citetalias{CL13}. While O/Mg and Al/Mg don't vary much, Si, Ca, and Ni yields decrease with respect to Mg between 8 and 30 $\msun$. Ni/Mg yields drop the most but are shallower than \citetalias{CL13}, suggesting a smaller IMF induced abundance change than that seen in Figure~\ref{fig:medDiffsIMF}. We have only chosen to plot a few select elements in these figures, but integrated net yields for all APOGEE elements can be found in Tables~\ref{tab:CL13_yield} (\citetalias{CL13}) and~\ref{tab:LC18_yields} (\citetalias{LC18}).

\begin{figure}[!htb]
 \begin{centering}
 \includegraphics[width=1\columnwidth, angle=0]{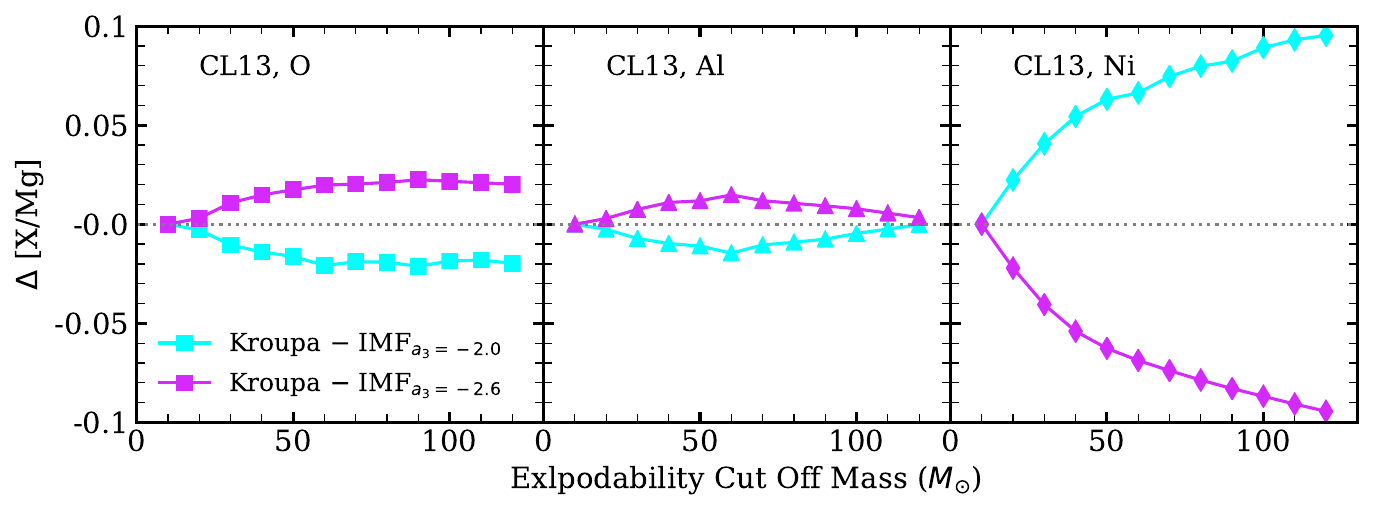}
 \caption{Theoretical $\xmg$ abundance changes between the standard Kroupa IMF ($a_3 = -2.3$) and a Kroupa IMF with an altered high mass slope as a function of changing explodability cutoff mass for O (left, squares), Al (center, triangles), and Ni (right, diamonds). Abundance changes for $a_3 = -2.0$ are shown in cyan and for $a_3 = -2.6$ are shown in magenta. Net yields from \citetalias{CL13} are used..}
 \label{fig:ApC_mass}
 \end{centering}
\end{figure}

The predicted abundance variability with changing IMF discussed in Section~\ref{sec:IMFs} employs net yields for $\feh=0$ and explodes stars up to a birth mass of $30\msun$. These choices impact the scale of the theoretical $\Delta \xmg$ values. To explore the full range of possible $\Delta \xmg$, we repeat our calculations for different explodability cutoffs and metallicities. Figure~\ref{fig:ApC_mass} plots the $\Delta \xmg$ values for the $a_3 = -2.0$ and $a_3 = -2.6$ cases as a we change the cutoff mass for O, Al, and Ni for the \citetalias{CL13} yields. A given mass on the $x$-axis indicates that all stars with progenitor masses $\leq M$ explode and contribute to nucleosynthesis while those above do not. The $\Delta$ [O/Mg] and $\Delta$ [Al/Mg] values remain small ($<0.03$), with the $\Delta$ [O/Mg] increasing to a plateau around $\pm 0.025$ and $\Delta$ [Al/Mg] peaking near $\pm 0.02$ for a cutoff mass of $\sim 60 \msun$. The $\Delta$ [Ni/Mg] values are much more dependent on the cutoff mass, growing to $\sim \pm 0.1$ as we include stars approaching $120\msun$. This limit gives an upper bound on the possible observable abundance changes induced by a changing IMF high mass slope. We also investigate how changing the metallicity affects the $\Delta \xmg$ values for LC18 yield, which are calculated for $\feh= -3.0, -2.0, -1.0, 0.0$. We find that the differences as a function of metallicity are smaller than those caused by the explosion mass cutoff, so we do not report them here. We will explore the impact of more complex explodability landscapes \citep{pejcha2015, tuguldur} in future work.

\begin{deluxetable*}{c|ccc|ccc}[h]
\tablecaption{VICE integrated yields (left) and [X/Mg] abundances (right) for net CL13 yields with [Fe/H] = 0.0, exploding stars up to 30 M$_{\odot}$. All APOGEE elements but Ce are reported for all three IMFs investigated above. \label{tab:CL13_yield}}
\tablehead{
\multicolumn{1}{c}{} & \multicolumn{3}{c|}{Yield ($\msunvice$)} & \multicolumn{3}{c}{[X/Mg]} \\
\cline{2-4}
\colhead{} & \colhead{$a_3 = -2.0$} & \colhead{Kroupa} & \multicolumn{1}{c|}{$a_3 = -2.6$} & \colhead{$a_3= -2.0$} & \colhead{Kroupa} & \colhead{$a_3= -2.6$}
}
\startdata
Fe & 1.09E-03 & 6.85E-04 & 3.48E-04 & -0.015 & 0.017 & 0.049 \\
O & 8.22E-03 & 4.66E-03 & 2.15E-03 & 0.214 & 0.204 & 0.193 \\
Na & 5.47E-05 & 3.10E-05 & 1.43E-05 & 0.329 & 0.318 & 0.307 \\
Al & 7.86E-05 & 4.50E-05 & 2.09E-05 & 0.208 & 0.201 & 0.195 \\
Si & 1.09E-03 & 6.51E-04 & 3.17E-04 & 0.273 & 0.285 & 0.297 \\
P & 5.59E-06 & 3.29E-06 & 1.58E-06 & 0.039 & 0.045 & 0.052 \\
S & 4.12E-04 & 2.47E-04 & 1.20E-04 & 0.181 & 0.194 & 0.208 \\
K & 4.78E-07 & 2.81E-07 & 1.35E-07 & -0.750 & -0.744 & -0.738 \\
Ca & 5.30E-05 & 3.19E-05 & 1.56E-05 & -0.025 & -0.010 & 0.004 \\
V & 1.94E-07 & 1.18E-07 & 5.89E-08 & -0.157 & -0.136 & -0.113 \\
Cr & 1.47E-05 & 8.93E-06 & 4.42E-06 & 0.005 & 0.023 & 0.044 \\
Mn & 8.66E-06 & 5.29E-06 & 2.65E-06 & -0.039 & -0.018 & 0.007 \\
Co & 3.65E-06 & 2.18E-06 & 1.08E-06 & -0.004 & 0.008 & 0.029 \\
Ni & 1.09E-04 & 7.08E-05 & 3.73E-05 & 0.243 & 0.291 & 0.338 \\
Cu & 3.12E-07 & 1.75E-07 & 8.02E-08 & -0.306 & -0.320 & -0.334 \\
\enddata
\end{deluxetable*}

\begin{deluxetable*}{c|ccc|ccc}[h]
\tablecaption{Same as Table~\ref{tab:CL13_yield}, but for \citetalias{LC18} with net explosive yields for stars up to $25 \msun$. All APOGEE elements are reported. \label{tab:LC18_yields}}
\tablehead{
\multicolumn{1}{c}{} & \multicolumn{3}{c|}{Yield ($\msunvice$)} & \multicolumn{3}{c}{[X/Mg]} \\
\colhead{} & \colhead{$a_3 = -2.0$} & \colhead{Kroupa} & \multicolumn{1}{c|}{$a_3 = -2.6$} & \colhead{$a_3= -2.0$} & \colhead{Kroupa} & \colhead{$a_3= -2.6$}
}
\startdata
Fe & 7.24E-04 & 4.57E-04 & 2.34E-04 & -0.103 & -0.077 & -0.050 \\
O & 5.96E-03 & 3.54E-03 & 1.71E-03 & 0.166 & 0.166 & 0.166 \\
Na & 3.65E-05 & 2.13E-05 & 1.01E-05 & 0.245 & 0.237 & 0.229 \\
Al & 4.94E-05 & 2.93E-05 & 1.41E-05 & 0.098 & 0.097 & 0.097 \\
Si & 8.51E-04 & 5.27E-04 & 2.65E-04 & 0.257 & 0.275 & 0.293 \\
P & 6.73E-06 & 4.11E-06 & 2.04E-06 & 0.212 & 0.224 & 0.237 \\
S & 3.14E-04 & 1.95E-04 & 9.89E-05 & 0.155 & 0.175 & 0.196 \\
K & 7.36E-07 & 4.48E-07 & 2.22E-07 & -0.470 & -0.460 & -0.449 \\
Ca & 4.20E-05 & 2.63E-05 & 1.34E-05 & -0.034 & -0.011 & 0.012 \\
V & 1.20E-07 & 7.50E-08 & 3.80E-08 & -0.273 & -0.252 & -0.230 \\
Cr & 8.11E-06 & 5.09E-06 & 2.60E-06 & -0.163 & -0.139 & -0.115 \\
Mn & 4.50E-06 & 2.81E-06 & 1.43E-06 & -0.232 & -0.210 & -0.188 \\
Co & 3.14E-06 & 1.92E-06 & 9.52E-07 & 0.022 & 0.034 & 0.046 \\
Ni & 6.67E-05 & 4.21E-05 & 2.16E-05 & 0.120 & 0.147 & 0.173 \\
Cu & 7.45E-07 & 4.36E-07 & 2.07E-07 & 0.164 & 0.157 & 0.150 \\
Ce & 1.71E-10 & 1.00E-10 & 4.76E-11 & -1.209 & -1.215 & -1.222 \\
\enddata
\end{deluxetable*}

\bibliography{bibliography} 
\bibliographystyle{aasjournal}

\end{document}